\documentclass[smallextended]{svjour3}       
\smartqed  
\usepackage{graphicx}

\usepackage{amsmath}
\usepackage{amsfonts}
\usepackage{amssymb}
\usepackage{mathrsfs}
\usepackage{bm}
\usepackage{graphicx}
\usepackage{xcolor}
\usepackage{epsfig}
\usepackage{setspace}
\usepackage{cite}
\usepackage{pdfpages}
\usepackage{hyperref}
\usepackage{afterpage}
\usepackage[toc,page]{appendix}
\usepackage{cancel}
\usepackage{array,multirow,makecell}

 \newcommand{\be}{\begin{equation}}
 \newcommand{\ee}{\end{equation}}
 \newcommand{\gm}

\begin{document}

\title{Fluid descriptions of quantum  plasmas}


\author{Giovanni Manfredi \and Paul-Antoine Hervieux \and J\'er\^ome Hurst
}


\institute{G. Manfredi
\at
              Universit\'e de Strasbourg, CNRS, Institut de Physique et Chimie des Mat\'eriaux de Strasbourg, UMR 7504, F-67000 Strasbourg, France \\
              \email{giovanni.manfredi@ipcms.unistra.fr}
           \and
           P.-A. Hervieux \at
              Universit\'e de Strasbourg, CNRS, Institut de Physique et Chimie des Mat\'eriaux de Strasbourg, UMR 7504, F-67000 Strasbourg, France
           \and
           J. Hurst
           \at  Universit\'e de Strasbourg, CNRS, Institut de Physique et Chimie des Mat\'eriaux de Strasbourg, UMR 7504, F-67000 Strasbourg, France
}

\date{Received: date / Accepted: date}

\maketitle

\def\mstar{m_{\displaystyle *}}
\def\mstar{m}

\begin{abstract}
Quantum fluid (or hydrodynamic) models provide an attractive alternative for the modeling and simulation of the electron dynamics in nano-scale objects.
Compared to more standard approaches, such as density functional theory or phase-space methods based on Wigner functions, fluid models require the solution of a small number of equations in ordinary space, implying a lesser computational cost. They are therefore well suited to study systems composed of a very large number of particles, such as large metallic nano-objects. They can be generalized to include the spin degrees of freedom, as well as semirelativistic effects such as the spin-orbit coupling. Here, we review the basic properties, advantages and limitations of quantum fluid models, and provide some  examples of their applications.

\end{abstract}


\section{Introduction -- Fluid models for classical and quantum plasmas}
\label{sec:intro}

Recent decades have witnessed a remarkable surge of interest for the electronic properties of nano-scale objects, particularly when excited by electromagnetic radiation \cite{Voisin2000,Bigot2000,Manfredi2018SPIE,Maniyara2019}. This is a very vast domain of research that encompasses all sorts of nano-objects (metallic films and nanoparticles, carbon nanotubes, semiconductor quantum dots \dots), new materials like graphene, as well as meta-materials whose structure can be engineered so as to display some particular electromagnetic properties. Potential applications are impressive, and range from high-performance computing (efficient storage and transfer of information), to nanoplasmonics  \cite{Stockman2011,Moreau2012} (optical filters, waveguides), and even to the medical sciences  (biomedical tests and sensors) \cite{Hainfeld2004,TatsuroEndo2006}.

The electronic  response of such systems (often in the out-of-equilibrium and nonlinear regimes) can be assimilated to that of a one-component quantum plasma and may be treated at different levels of approximations. For systems containing many interacting electrons, condensed-matter physicists and quantum chemists have developed various theoretical strategies, such as the time-dependent density functional theory (TDDFT) and time-dependent Hartree-Fock (TDHF) methods. TDDFT and TDHF are wavefunction-based methods, but they can be reformulated in the phase-space language familiar to plasma physicists by making use of Wigner functions. In a recent work \cite{Manfredi2019}, we reviewed the use of phase-space methods  for applications to condensed matter and nanophysics.  The first chapters of that work also discuss some general properties of solid-state plasmas.

Both approaches (wavefunction and phase space) are rather costly in terms of run-time and memory storage, especially for systems containing thousands of electrons. A possible, less costly alternative is provided by fluid models, which can be derived from the corresponding kinetic equations (Wigner for a fully quantum approach, or Vlasov for semiclassical modeling) by taking velocity moments of the distribution function \cite{Manfredi2001}. Although some information is inevitably lost in this procedure, one can expect the fluid approach to be accurate enough to provide key insights on the underlying physical mechanisms, while at the same time remaining affordable in terms of computational cost.
Like their classical counterparts, the validity of quantum fluid models is restricted to long wave lengths compared to the interparticle distance \cite{Khan2014}. However, they can deal with nonlinear effects (large excitations), quantum effects (tunnelling), Coulomb exchange (an effect related to the Pauli exclusion principle), and electron-electron correlations.

In the fluid approach, the electron dynamics is described by a set of  hydrodynamic equations (continuity, momentum balance, energy balance) that include quantum effects via the so-called Bohm potential. A considerable gain in computing time can be expected in comparison to Wigner or TDDFT simulations: indeed, TDDFT methods must solve $N \gg 1$ Schr\"odinger-like equations, while the phase space approach doubles the number of independent variables (positions and velocities). Hydrodynamic models were used in the past in condensed-matter physics, particularly for semiconductors \cite{Muller2004}  and, to a lesser extent, metal clusters \cite{Domps1998,Banerjee2000}.
More recently, the quantum fluid description has been extended to include another important property of the electron, namely its spin \cite{Brodin2007}. The resulting fluid equations are much more involved that their spinless counterparts, but may find useful applications, in condensed-matter physics, to the emerging field of spintronics \cite{Hirohata2020}, and in plasma physics to the study of highly polarized electron beams \cite{Wu2019,Wu2020}.
Quantum fluid theory finds further applications to the dense plasmas produced in the interaction of solid targets with intense laser beams \cite{Kremp2001}, to warm dense matter experiments \cite{Dornheim2018}, {\gm to quantum nanoplasmonics \cite{Ciraci2013,Ciraci2016,Ciraci2021,Toscano2015}}, and to compact astrophysical objects such as white dwarf stars \cite{Uzdensky2014}.

The purpose of this short review is to present the results on quantum fluid models obtained in our research group at the University of Strasbourg during the last two decades. The bibliography on quantum plasmas has grown immensely during this  period and it is hardly possible to do justice to all works on this topic. A recent paper \cite{Bonitz2019} has tried to summarize the status of present quantum fluid theory, as well as the prospects for future developments.

The present work is organized as follows. In section \ref{Fluid models} we derive the fluid equations in the simplest case and describe their properties. In section \ref{sec:validity} we discuss the range of validity of the fluid approach. In particular, the closure relations are analyzed by comparing the fluid and kinetic dispersion relations in two regimes of interest: fast Langmuir waves and slow ion acoustic waves. Section \ref{metal_nanoshells} contains a practical example of application of quantum fluid theory to plasmonic oscillations in a metallic nanoshell.
Section \ref{Fluid model with spin effects} generalizes the fluid equations to take into account the electron spin. More advanced semirelativistic effects, such as the spin-orbit coupling, are also considered.
In section \ref{sec:variational} we discuss a variational formulation of the quantum fluid equations through an appropriate Lagrangian.
The variational approach enables us to reduce the full fluid description to a set of a few ordinary differential equations, which can then be solved either analytically or numerically with little computational cost. Some applications to electronic modes in semiconductor quantum wells are also illustrated.

\section{Quantum fluid models without spin}\label{Fluid models}

For a classical collisionless plasma, the electron dynamics is fully described by a probability distribution in the phase space $f(\bm{r},\bm{p}, t )$, evolving in time according to the Vlasov equation. For a quantum electron gas, the corresponding statistical tool is the density matrix:
\begin{align}
\rho(\bm{r},\bm{r}') = \sum_{\alpha =1}^{N} p_{\alpha} \Psi_{\alpha}^{\displaystyle *}(\bm{r})  \Psi_{\alpha}(\bm{r}') ,
\label{density matrix quantum mixture of state}
\end{align}
where $p_{\alpha}$ is the probability to be in the state $\Psi_{\alpha}$.
In order to make contact with classical plasma physics, it is useful to introduce the  Wigner distribution function, defined as
{\gm
\begin{align}
f\left(\bm{r},\bm{p}, t \right) =
\frac{1 }{\left(2 \pi \hbar\right)^{3}}   \int d \bm{\lambda} \exp \left( \frac{i \bm{p} \cdot \bm{\lambda} }{\hbar} \right)\rho\left(\bm{r} + \frac{\bm{\lambda}}{2}, \bm{r} - \frac{\bm{\lambda}}{2}, t \right).
\label{one body dis function exp}
\end{align}
}
The Wigner function evolves in time according to the following Wigner equation
\be
\frac{\partial f}{\partial t} + \frac{\bm{p}}{m}\cdot \bm{\nabla}f =
\displaystyle
\frac{ie}{\hbar}\frac{1}{\left( 2 \pi \hbar \right)^{3}} \int d\bm{\lambda} d\bm{p'} e^{\frac{ i \left(\bm{p}- \bm{p'} \right) \cdot \bm{\lambda}}{\hbar}} \left[ V( \bm{r}_{+}) - V( \bm{r}_{-}) \right] f(\bm{r},\bm{p'},t). \label{wigner poisson equ}
\ee
where $e>0$ and $m$ are the electron charge and mass respectively, the indices $\bm{\pm}$ denote the shifted positions $\bm{r}_{\pm} = \bm{r} \pm \bm{\lambda} /2$, and the potential $V(\bm{r},t)$ is either an external potential or a self-consistent potential obtained from Poisson's equation (or the sum of both).

The Wigner approach allows one to recast the dynamics of a quantum electron gas in the familiar formalism of the classical phase space. But it should not be forgotten that the Wigner function, although real, can take negative values, and is therefore not a true probability distribution like its classical counterpart. Two comprehensive reviews of the Wigner phase-space methods applied to the description of quantum plasmas were published recently in this journal \cite{Manfredi2019,Melrose2020}, to which we refer the reader for further details.

Fluid, or hydrodynamic, models are usually obtained by taking  moments of the relevant phase-space distribution (Vlasov or Wigner). Thus, the $k$-th order moment is defined as:
\be
\textsf{m}_k (\bm{r}, t) \equiv \int f(\bm{r},\bm{p}, t)\, \underbrace{\bm{p} \otimes  \bm{p} \dots \otimes \bm{p}}_{k \, \rm times} \, d\bm{p},
\ee
where the symbol $\otimes$ denotes the tensor product. Clearly, $\textsf{m}_k$ is a tensor of rank $k$.
Each moment obeys an evolution equation, which can be interpreted as a conservation law for a given physical quantity (mass, momentum, energy, etc\dots). The evolution equation for the moment of order $k$ generally depends on the equation for the moment of order $k+1$, thus forming an infinite hierarchy of equations. Hence, some further assumptions are needed in order to close this infinite system. Here, we will essentially consider two-moment fluid systems, based on the evolution of the electron density $n(\bm{r}, t)$ (zeroth order moment):
\begin{align}
\int f d\bm{p} =  \sum_{\alpha =1}^{N} p_\alpha \Psi _{\alpha}^{*}\left(\bm{r}, t \right)\Psi _{\alpha}\left(\bm{r}, t \right) \equiv n\left(\bm{r}, t \right),
\end{align}
and the electron current $\bm{j}(\bm{r}, t)$ (first order moment):
\begin{align}
\int \bm{p} f d\bm{p} = \frac{i\hbar}{2}  \sum_{\alpha =1}^{N} p_\alpha \left[ \Psi _{\alpha}\left(\bm{r}, t \right) \bm{\nabla} \Psi _{\alpha}^{*}\left(\bm{r}, t \right) -  \Psi _{\alpha}^{*}\left(\bm{r}, t \right) \bm{\nabla} \Psi _{\alpha}\left(\bm{r}, t \right)\right] \equiv \bm{j}\left(\bm{r}, t \right).
\end{align}
{\gm A generalization of this procedure to higher order moments was also developed \cite{Haas2010}.}

By integrating the Wigner equation \eqref{wigner poisson equ} over the momentum variable, we obtain the continuity equation (conservation of  mass):
\begin{align}
\frac{\partial n}{\partial t} + \bm{\nabla}\cdot \left( n\bm{u} \right)=0,
\label{continuity equation}
\end{align}
where we  used the mean velocity $\bm{u}(\bm{r}, t ) \equiv \bm{j}(\bm{r}, t ) /n(\bm{r}, t )$.
Multiplying the Wigner equation by $\bm{p}$ and integrating yields the evolution equation for the mean velocity (momentum conservation law):
\begin{align}
\frac{\partial u_{i} }{\partial t} + u_{j} (\partial_{j} u_{i}) =- \frac{1}{nm} \partial_{j} \textsf{P}_{ij} + \frac{e}{m} \partial_{i} {V},
\label{Euler equation}
\end{align}
where we adopted the Einstein summation convention over repeated indices (this will be used systematically in the rest of this work). Note that Eq. \eqref{Euler equation} has the same form as the Euler equation in standard hydrodynamics.

The pressure tensor $\textsf{P}_{ij}$ is a second-order moment of the distribution function defined as
\begin{align}
\textsf{P}_{ij}&= \int w_{i} w_{j} fd\bm{p},
\label{pressure term second order moment}
\end{align}
where we separated the mean fluid velocity $\bm u$ from the velocity fluctuations $\bm{w} \equiv \bm{v} - \bm{u}$.
To obtain the evolution equation of the pressure tensor, one should take the second order moment of the Wigner equations, which would contain the third-order moment (energy flux), and so on and so forth. As we are aiming at a two-moment fluid model, we should try to obtain some closure relation that allows us to express $\textsf{P}_{ij}$ as a function of the lower order moments $n$ and $\bm{u}$.

A simple closure can be achieved by  writing the pressure tensor in terms of the electronic wave functions. Using Eqs. \eqref{one body dis function exp} and \eqref{pressure term second order moment} and after some algebra, one is able to write the pressure tensor as follows:
\begin{align}
\textsf{P}_{ij} =& \frac{\hbar ^{2}}{4m}  \sum _{\alpha} p_{\alpha}  \left[  \left( \partial_{i} \Psi _{\alpha} ^{*} \right) \left( \partial_{j} \Psi _{\alpha} \right) + \left( \partial_{j} \Psi _{\alpha} ^{*} \right) \left( \partial_{i} \Psi _{\alpha} \right) -  \Psi _{\alpha} ^{*}\left[ \partial_{i} \left( \partial_{j} \Psi _{\alpha} \right)\right] -  \Psi _{\alpha} \left[ \partial_{i} \left( \partial_{j} \Psi _{\alpha}^{*} \right)\right] \right] \nonumber \\
&~  + \frac{\hbar ^{2}}{4mn} \left[ \sum _{\alpha} p_{\alpha} \left[ \Psi _{\alpha} ^{*} \left(\partial_{i} \Psi _{\alpha} \right) -  \Psi _{\alpha}  \left(\partial_{i} \Psi _{\alpha}^{*} \right) \right] \left[ \Psi _{\alpha} ^{*} \left(\partial_{j} \Psi _{\alpha} \right) -  \Psi _{\alpha}  \left(\partial_{j} \Psi _{\alpha}^{*} \right) \right] \right]^{2}. \label{pression wave function}
\end{align}
In order to interpret this pressure tensor, we shall use the Madelung decomposition \cite{Madelung1927}:
\begin{align}
\Psi _{\alpha}(\bm{r},t) &= A_{\alpha} (\bm{r},t) \exp \left(\frac{i S_{\alpha} (\bm{r},t) }{\hbar} \right),
\label{madelung decomposition}
\end{align}
where $A_{\alpha}(\bm{r},t)$ is the amplitude of the wave function and $S_{\alpha}(\bm{r},t)$ its phase, both being real functions.

The individual particle density and velocity \emph{for each wave function} are defined as:
\begin{align}
n_{\alpha}(\bm{r},t) &= A_{\alpha}^{2}(\bm{r},t),~~~~\bm{u}_{\alpha}(\bm{r},t)= \frac{1}{m} \bm{\nabla} S_{\alpha}(\bm{r},t).
\label{individual density velocity madelung}
\end{align}
The connection with the global fluid variables $n$ and $\bm{u}$, derived from the Wigner approach, is made by taking the statistical average of Eqs.  \eqref{individual density velocity madelung}, to obtain
\begin{align}
n &=  \sum _{\alpha=1}^{N} p_{\alpha} n_{\alpha} ~~~~\textrm{and}~~~~ \langle \bm{u}_\alpha \rangle \equiv \bm{u} = \frac{1}{n} \sum _{\alpha=1}^{N} p_{\alpha} n_{\alpha} \bm{u}_{\alpha}.
\label{density velocity madelung}
\end{align}
Then using, Eqs. \eqref{madelung decomposition}-\eqref{density velocity madelung}, one obtains a simpler form for the pressure tensor:
 \begin{align}
\textsf{P}_{ij} &= m n\left( \langle u_{i}u_{j} \rangle - \langle u_{i} \rangle \langle u_{j} \rangle \right)  \nonumber \\
& + \frac{\hbar ^{2}}{2m} \sum _{\alpha} p_{\alpha} \left[ \left( \partial_{i} \sqrt{n_{\alpha}}\right) \left( \partial_{j} \sqrt{n_{\alpha}}\right) -  \sqrt{n_{\alpha}} \left( \partial_{i} \left(\partial_{j} \sqrt{n_{\alpha}} \right) \right] \right] \equiv \textsf{P}^C_{ij}+\textsf{P}^Q_{ij}.
\label{pression c + q}
\end{align}

The first term of the right-hand side ($\textsf{P}^C_{ij}$) has the form of a velocity dispersion, i.e., the same form as the standard definition of the pressure tensor in an ordinary classical gas.
It is a statistical term which disappears for a pure quantum state, i.e., when only one quantum state is occupied in Eq. \eqref{density matrix quantum mixture of state}.
Hence, we can use our knowledge of classical fluid dynamics to obtain appropriate closures for this  ``classical" pressure term. For instance, for an isotropic Maxwell-Boltzmann equilibrium at constant temperature $T_e$ (isothermal transformation), one would have: $\textsf{P}^C_{ij}=  \delta_{ij} P[n]$, with $P[n]= n k_B T_e$, where $k_B$ is the Boltzmann constant and $\delta_{ij} $ is the Kronecker delta.
For an ideal and fully degenerate electron gas, this ``classical" and isotropic pressure would coincide with the Fermi degeneracy pressure \cite{Ashcroft2002}:
\begin{align}
 P [n]= \frac{\left(3 \pi^{2}\right)^{2/3}\hbar^{2}}{5m} n^{5/3}.
\label{pressure classical}
\end{align}
The above examples are simple instances of equations of state (EOS).

The second term of the right side is proportional to Planck's constant $\hbar$, hinting at its quantum origin. Indeed, this quantum pressure term $\textsf{P}^{Q}_{ij} $ has no classical counterpart and subsists even when only one state is occupied in Eq. \eqref{density matrix quantum mixture of state}, i.e. for a pure quantum state. Physically, it originates from the Heisenberg uncertainty principle, which forbids that a quantum particle possesses a definite velocity, except if its position is completely delocalized. For instance, it is in virtue of this ``quantum" pressure that the ground state of a quantum harmonic oscillator has a finite velocity dispersion, unlike a classical oscillator.

An exact closure for the quantum pressure can be found under the (rather restrictive)  assumption that the amplitudes of all the wave functions are identical, i.e. $n_{\alpha}(\bm{r},t) = n(\bm{r},t), \, \forall \alpha $  \cite{Manfredi2005}. In this case one obtains:
\begin{align}
\textsf{P}^{Q}_{ij} &=  \frac{\hbar ^{2}}{2m} \left[ \left( \partial_{i} \sqrt{n} \right)\left( \partial_{j} \sqrt{n} \right) - \sqrt{n}\left[\partial_{i} \left( \partial_{j} \sqrt{n}\right) \right]\right].
\label{pression Q n}
\end{align}
To simplify even more, one can assume the isotropy of the pressure, i.e. $\textsf{P}^{Q}_{ij}  = P^{Q} \, \delta_{ij}$, yielding
\be
P^{Q}[n] = \frac{\hbar ^{2}}{2m} \left[ ( \bm{\nabla} \sqrt{n})^2 - \sqrt{n} \,\bm{\nabla}^2 (\sqrt{n})  \right]
\label{pression Q n_iso}
\ee
where we have indeed expressed $P^{Q}$ in terms of the electron density and its derivatives.
This completes the closure procedure.
We also note that the quantum pressure \eqref{pression Q n_iso} can be rewritten in the form of a potential, by noting that
\be
\frac{\bm{\nabla} P^Q }{m} = \bm{\nabla} V_B,
\ee
where $V_B$ is the Bohm potential \cite{Bohm1952}, defined as:
\begin{align}
V_B= -\frac{\hbar^{2}}{2m} \frac{\bm{\nabla}^{2} \sqrt{n} }{ \sqrt{n}}.
 \label{bohm potential}
 \end{align}

 {\gm
 However, we note that the hypothesis of identical amplitudes for all the wave functions, although certainly sufficient to obtain an {\em exact} closure of the quantum pressure term, is by no means necessary for {\em approximate} closures. Indeed, the Bohm potential is in reality a quantum correction to the electron kinetic energy \cite{Michta2015}, as was already noticed by von  Weizs\"acker many years ago \cite{Weizsacker1935}. It can be derived from the standard Thomas-Fermi theory including gradient corrections, and it appears as the first gradient correction beyond the local density approximation. As a gradient correction, it disappears for a homogeneous density profile and is small for weak spatial modulations (long wave lengths). This suggests that the closure \eqref{pression Q n_iso} is approximately correct for density gradients that are not too strong. A simple dimensional analysis shows that, for an electron gas at zero temperature, the scale length beyond which this closure is acceptable must be of the order of the Thomas-Fermi screening length $\lambda_{TF} = \sqrt{2 \varepsilon_0 E_F /(e^2 n_0)}$, where $e$ is the absolute electron charge, $\varepsilon_0$ is the dielectric constant in vacuum, $n_0$ is a reference number density, and $E_F=  \frac{\hbar}{2m} \left( 3 \pi^{2} n_0\right)^{2/3}$ is the Fermi energy.
 }

Then, using the closure relations for the classical pressure \eqref{pressure classical} and the quantum pressure \eqref{pression Q n_iso}, one can rewrite the fluid equations \eqref{continuity equation} and \eqref{Euler equation} in the closed form:
\begin{align}
\begin{array}{lcl}
\displaystyle
\displaystyle\frac{\partial n}{\partial t} + \bm{\nabla} \cdot \left(n\bm{u}\right)  = 0,   \\ \\
\displaystyle\frac{\partial u_{i} }{\partial t} + u_{j} (\partial_{j} u_{i}) = \frac{\hbar^{2}}{2m^{2}} \partial_{i} \left[\frac{\bm{\nabla}^{2} \sqrt{n} }{ \sqrt{n}} \right] -\frac{1}{nm}\partial_{i} P + \frac{e}{m} \partial_{i} {V}.
\end{array}
\label{fluid_equation_closed_charge}
\end{align}

The above quantum fluid equations \eqref{fluid_equation_closed_charge} are usually referred to as the \emph{quantum hydrodynamic (QHD)} model. The potential $V(\bm{r},t)$ can often be written as the sum of an external potential $V_{ext}$ and a self-consistent Hartree potential $V_H$, which is a solution of the Poisson equation
 \be
 \bm{\nabla}^2 V_H= e n/\varepsilon_0.
 \label{eq:poisson}
 \ee
 Equations \eqref{fluid_equation_closed_charge}-\eqref{eq:poisson}  constitute a useful semiclassical mean-field model to treat the electron dynamics of a degenerate electron gas, simpler to implement numerically than the corresponding Wigner equation. They were used, for instance, to study the nonlinear  electrons dynamics in metallic films \cite{Crouseilles2008} or, more recently, monopole plasmon oscillations in a $\rm C_{60}$ molecule \cite{Tanjia2018} and in a metallic nanoshell \cite{Tanjia2018}.

Further, the total potential $V$ in Eq. \eqref{fluid_equation_closed_charge} can be augmented to include non-ideal effects such as electronic correlations and the exchange interaction.
 {\gm The exchange interaction stems from the fact that the many-body wave function of a system of uncorrelated fermions is not simply the product of single-particle wave functions, but rather has the form of a Slater determinant. This guarantees that the many-body wave function is antisymmetric (changes sign when two particles are interchanged), as it should be for fermions \cite{Manfredi2019}. }
Both exchange and correlations can be included in the fluid models, following the approach of density functional theory (DFT), by defining appropriate potentials $V_C[n]$ and  $V_X[n]$ that depend functionally on the electron density. The simplest choice is the so-called local density approximation (LDA) \cite{Kohn1965}, whereby the exchange and the correlation functionals depend locally on the electron density.
For instance, the LDA approximation for the exchange potential is:
\be
V_{X} [n]= -\frac{e^{2}}{4 \pi\epsilon_{0}}\left( \frac{3}{\pi} \right)^{1/3} n^{1/3} .
\label{eq:xlda}
\ee

{\gm More sophisticated functionals have been designed to describe electron correlations in the transition region between the bulk and the outer surface of a nano-object \cite{Armiento2005}.}
Many other approximate functionals have been developed over the years (such as the generalized gradient approximation, GGA), making DFT methods a cornerstone of computational materials science and theoretical chemistry \cite{Jones2015}.
{\gm A recent  benchmarking study compared the performances of  various commonly used exchange-correlation functionals regarding their ability to describe electronic systems under external harmonic perturbations with different amplitudes and wave-numbers \cite{Moldabekov2021b}.
These functionals can be used in fluid models to extend their validity beyond the mean-field approximation.}

{\gm Alternatively, the exchange interaction can be taken into account in a phase-space description, such as the Wigner or Vlasov equations. As the exchange is a two-body effect (beyond mean field), one must include corrections brought about by the two-body distribution function to derive a suitable kinetic equation \cite{Zamanian2013}. Next, one can follow the same moment-taking procedure as above to obtain a set of fluid equations that include the electron exchange interaction, as was done very recently in Ref. \cite{Haas2021}.
}

Finally, we  note that the fluid equations \eqref{fluid_equation_closed_charge} can be viewed as a time-dependent generalization of the early Thomas-Fermi theory of the atomic electron gas \cite{Thomas1926}. Indeed, taking $\partial/\partial t =0$ and $\bm{u}=0$ everywhere, and neglecting the quantum term (i.e., the Bohm potential), one gets:
\[
- \frac{\nabla P}{n} +e\nabla V_H + e\nabla V_{ext} = 0.
\]
Using Eq. \eqref{pressure classical}, and defining the chemical potential $\mu$ as an integration constant, we obtain
\be
- \frac{\hbar^2}{2m} (3\pi^2 n)^{2/3} +e V_H[n] + e V_{ext} = \mu ,
\ee
which is  an integral equation for the ground-state density $n$ and is identical to  the standard Thomas-Fermi equation \cite{Michta2015}.


\section{Validity of quantum fluid models}\label{sec:validity}

\subsection{Closure relations}
As we have seen in the preceding section, the fluid equations need two further closure relations in order to form closed system -- one for the classical pressure and one for the quantum pressure (or Bohm potential). The validity of the model thus relies on the accuracy of such closure  hypotheses. Several recent works have analyzed this issue and suggested procedures to improve the simple closure relations mentioned above \cite{Haas2015,Moldabekov2018,Vladimirov2011}.

In this section, we will investigate the closure relations by computing the linear dispersion relations for the fluid model and the corresponding kinetic one (Wigner-Poisson) and comparing the two results. This will be done for two extreme cases of high-frequency Langmuir waves (plasmons) and low-frequency ion acoustic waves. It will appear clearly that the closure relations are not universal, but have rather to be adapted to the physical situation under study.

To fix the ideas, we choose a polytropic EOS for the classical pressure:
\be
P = P_0  \left( n \over n_0 \right)^\gamma ,
\label{eq:polytropic}
\ee
where $n_0$ is an equilibrium density and $P_0=n_0 k_B T_e$ is the equilibrium pressure. By taking small fluctuations around the equilibrium ($n=n_0+n_1 + \dots$ and $P=P_0+P_1 + \dots$), one can write: $P_1 = \gamma k_B T_e n_1$. Thus, for $\gamma=1$ one  recovers the isothermal case for an ideal classical gas and for $\gamma=5/3$ the ideal Fermi degeneracy pressure [see Eq. \eqref{pressure classical}].

For the quantum pressure, we stick with the isotropic expression \eqref{pression Q n_iso} -- or equivalently the Bohm potential \eqref{bohm potential} -- but  multiply it by a (yet arbitrary) factor $\zeta>0$. It will be apparent in the forthcoming discussion that this factor need not be equal to unity.

In summary, we have constructed a two-parameter family of closure relations, spanned by the parameters $(\gamma, \zeta)$. In the next subsections, we will determine their values in two specific cases of linear wave propagation.

\subsection{Linear dispersion relation: High-frequency Langmuir modes} \label{sec:langmuir}

\paragraph{Kinetic theory}
The kinetic dispersion relation for longitudinal Langmuir waves (plasmons) is obtained by linearizing the electron Wigner equation \eqref{wigner poisson equ}
together with the Poisson equation for both electrons and fixed ions:
\be
\bm{\nabla} V = e(n-n_0)/\varepsilon_0 .
\label{eq:poisson-ion}
\ee
The ions are supposed to be fixed and constitute a homogeneous neutralizing background with density $n_0$, equal to the equilibrium electron density. The linearization procedure yields the following longitudinal dielectric function \cite{Klimontovich1960,Tyshetskiy2011}
\begin{align}
\epsilon(\omega,\bm{k}) = 1+ \frac{\omega_{p}^{2}m}{\hbar \bm{k}^{2} n_{0}} \int d\bm{v} \frac{f^{(0)}\left(\bm{v}+\hbar \bm{k}/2m\right)-f^{(0)}\left(\bm{v}-\hbar \bm{k}/2m \right)}{\omega - \bm{k} \cdot \bm{v}},
\label{longitudinal dielectric function charge}
\end{align}
where $f^{(0)}(\bm{v})$ is the equilibrium electron distribution function, $n_0=\int f^{(0)} d \bm{v}$ is the equilibrium density, and  $\omega_{p}=\sqrt{e^2 n_0/(m\varepsilon_0)}$ is the electron plasma frequency. Note that here we use the velocity $\bm{v} =\bm{p} /m$ instead of the momentum.

To simplify the analysis, we consider a one-dimensional system along the direction of propagation of the Langmuir waves, denoted $x$, and rename $v_x \to v$ and  $k_x \to k$ for the wave vector.
The dispersion relation $\omega(k)$ is obtained by setting  $\epsilon(\omega,k)=0$, where $\omega$ and $k$ are, respectively, the complex frequency and the wave vector of the excitation modes. Finding the complete dispersion relation is in general a challenging problem , because of the presence of a singularity in the denominator of Eq. \eqref{longitudinal dielectric function charge} {\gm (for a recent attempt at solving this problem rigorously, see \cite{Hamann2020})}.
The correct treatment of this singularity in the complex plane yields the imaginary part of the frequency, which represents Landau damping.

Here, we do not consider this effect and only focus on the real part of the frequency. Further, we make two assumptions, namely that (i) quantum effects are small and that (ii) the wavelength of the modes are large. The first assumption means that  $\hbar k / (2m) \ll v$. In this case, we can perform a Taylor expansion on $f^{(0)}\left(v \pm \hbar k/2m\right)$, yielding:
\begin{align}
f^{(0)}\left(v \pm \frac{\hbar k}{2m} \right) = f^{(0)}\left(v \right) \pm  f^{(0)'}\left(v \right) \frac{\hbar k}{2m} + \frac{1}{2!} f^{(0)''}\left(v \right) \left(\frac{\hbar k}{2m}\right)^{2} + \cdots~,
\end{align}
where the apex denotes differentiation with respect to $v$.
Substituting into the dielectric function, one obtains:
\begin{align}
\epsilon(\omega,k) = 1+ \frac{\omega_{p}^{2}}{ k n_{0}} \int dv \,\frac{f^{(0)'}\left(v\right)}{\omega - kv} +  \frac{\omega_{p}^{2}\hbar^{2}k}{ 24 m^{2} n_{0}} \int dv\, \frac{f^{(0)'''}\left(v\right)}{\omega - kv}+\cdots~.
\end{align}
As expected, if we set $\hbar=0$ in the above equation, we recover the dielectric function corresponding to the Vlasov-Poisson equations \cite{Lyu2014}.

The second assumption (long wavelengths), implies that $ k \ll \omega / v $, hence there is no singularity in the denominator: in this limit, the frequency is real and Landau damping vanishes. This is of course consistent with fluid models, which also contain no Landau damping.
By performing the expansion up to  third order in $k v / \omega$:
\begin{align}
\frac{1}{\omega - kv} =\frac{1}{\omega} + \frac{kv}{\omega^{2}} + \frac{k^{2}v^{2}}{\omega^{3}} + \cdots~,
\label{dev long wave length}
\end{align}
the dielectric function becomes:
\begin{align}
\epsilon(\omega,k)
=
1- \frac{\omega_{p}^{2}}{ \omega^2}
 - 3 \frac{k^2\omega_p^{2}}{\omega^4}   \langle v^2 \rangle
- \frac{\omega_{p}^{2}\hbar^{2}k^4}{ 2 m^{2}} \frac{1}{\omega^{4}} +\cdots~,
\label{eq:dielectric}
\end{align}
where we used the  equilibrium mean square velocity, defined as
\be
\langle v^2 \rangle = \frac{1}{n_0}\, \int f^{(0)} v^2 dv .
 \ee
 Keeping only terms with small wave vector $k$ (long wavelength), we obtain the dispersion relation
\begin{align}
\omega^{2} = \omega_{p}^{2} +3  k^{2} \langle v^2 \rangle + \frac{\hbar^{2} k^{4}}{4 m^{2}}+\cdots~.
\label{relation de dispersion kin}
\end{align}
This is a quantum extension of the well-known Bohm-Gross dispersion relation, which was found earlier by several authors \cite{Pines1961,Manfredi2005}.
Note that for a Maxwell-Boltzmann distribution $\langle v^2 \rangle =k_B T_e/m \equiv v_{th}^2$, whereas for a Fermi-Dirac at zero temperature (fully degenerate gas) $\langle v^2 \rangle =   k_B T_F/(5m)$, where $T_F = \frac{\hbar}{2m k_{{B}}} \left( 3 \pi^{2} n_0\right)^{2/3}$ is the Fermi temperature of the electron gas.

\paragraph{Fluid theory}
Next, we compute the fluid dispersion relation by linearizing the continuity and Euler equations \eqref{fluid_equation_closed_charge}, together with the Poisson equation for both electrons and fixed ions \eqref{eq:poisson-ion}.
We use the EOS \eqref{eq:polytropic} for the classical pressure and, as in the Wigner-Poisson case, restrict our analysis to one spatial dimension along $x$. We expand all quantities around the equilibrium $\{n_0,u_0=0,V_0=0\}$ by writing $n=n_0+n_1$, $u=u_0+u_1$, etc\dots, and neglect second order terms such as $n_1 u_1$. The first-order classical pressure reads as:  $P_1 = \gamma k_B T_e n_1$. Then, expressing all first-order quantities in term of plane waves
\be
n_1(x,t) = \bar{n}_1 \exp\left(-i\omega t + i k x  \right) ,
\label{eq:planewaves}
\ee
we arrive at the fluid dispersion relation:
\be
\omega^{2} = \omega_{p}^{2} + \gamma  k^{2} v_{th}^2 + \zeta \frac{\hbar^{2} k^{4}}{4 m^{2}}~.
\label{relation de dispersion fluid}
\ee

For a Maxwell-Boltzmann equilibrium ($\langle v^2 \rangle = v_{th}^2$), the kinetic \eqref{relation de dispersion kin} and fluid \eqref{relation de dispersion fluid} dispersion relations become identical, for small wave vectors, if one chooses $\gamma=3$ and $\zeta=1$.
Hence, comparison of the two models suffices to fully determine the closure relations, at least in the linear response regime.
These closures can be understood as follows. For a classical ideal gas, the polytropic exponent for an adiabatic and reversible (isentropic)  transformation is $\gamma = (d+2)/d$ where $d$ is the number of degrees of freedom (equal to the number of spatial dimensions for point-like particles). Then, one has to choose $d=1$ because Langmuir waves (plasmons) propagate fast in one direction, with no exchange of energy in the transverse plane, so that only one degree of freedom is effectively active.
The choice $\zeta=1$ amounts to choosing the quantum propagation velocity of a \emph{free} electron obeying the Schr\"odinger equation.  Again, this choice is dictated by the fast motion of Langmuir waves, which, to lowest order in $\hbar$, do not feel the Coulomb potential.

\subsection{Linear dispersion relation: Low-frequency ion acoustic modes}
Here, we discuss the linear response theory of ion acoustic waves
{\gm (phonons),
which are routinely observed in solid state plasmas \cite{Ma2015}.}
For such waves, the phase velocity is intermediate between the ion and the electron thermal speeds:
\be
v_{th,i} \ll {\omega \over k} \ll v_{th,e} ,
\label{eq:phasevel}
\ee
where $v_{th,i,e} = \sqrt{k_B T_{i,e}/m_{i,e}}$. It will also be assumed that the ions are classical and cold, i.e. $T_i \ll T_e$, while the electrons are quantum.
We shall follow the derivation detailed in   \cite{Haas2003} and \cite{Haas2015}.

\paragraph{Kinetic theory}
Under these hypotheses, and also assuming wave propagation along the $x$ direction only, the Wigner-Poisson dielectric constant reads as:
\be
\epsilon(\omega,k) = 1 - \frac{\omega_{pi}^2}{\omega^2} + \frac{\omega_{pe}^{2}m}{\hbar k^{2} n_{0}} \int d{v}\, \frac{f_e^{(0)}\left({v}+\hbar k/2m\right)-f_e^{(0)}\left({v}-\hbar {k}/2m \right)}{\omega - k  v},
\label{dielectric function acoustic}
\ee
where $\omega_{p\,i,e}^2 = e^2 n_0/(m_{i,e} \varepsilon_0)$.
The last term in Eq. \eqref{dielectric function acoustic} can be expanded in powers of $\hbar$, yielding:
\[
 \frac{\omega_{pe}^2}{kn_ 0}\int \frac{f_e^{(0)'}(v)}{\omega - k  v} dv +
 \frac{\hbar^2 \omega_{pe}^2 k}{24 n_ 0 m_e^2}\int \frac{f_e^{(0)'''}(v)}{\omega - k  v} dv + \dots.
\]
Because of the hypothesis \eqref{eq:phasevel}, we can neglect the frequency $\omega$ in the  above expressions. Taking a Maxwell-Boltzmann distribution
\be
f_e^{(0)} = \frac{n_0}{\sqrt{2\pi}\, v_{th,e}} \, \exp\left( -\frac{v^2}{2 v^2_{th,e}}\right) ,
\label{eq:maxwellian}
\ee
we can compute the intergals:
\[
\int \frac{f_e^{(0)'}(v)}{v} dv = -\frac{n_0}{v^2_{th,e}}\, , \hskip1cm
\int \frac{f_e^{(0)'''}(v)}{v} dv = \frac{2 n_0}{v^2_{th,e}} ,
\]
where we used $\langle v^2 \rangle = v^2_{th,e}$.

Inserting into the dielectric constant \eqref{dielectric function acoustic} and setting $\epsilon(\omega,k) = 0$, yields the kinetic dispersion relation
\be
\omega_{\rm kin}^2 = \frac{c_s^2 k^2}{1+\left(1-\frac{H^2}{12}\right) k^2 \lambda_D^2} ,
\label{eq:disp_relation_acoustic_kin}
\ee
where $\lambda_D=v_{th,e}/\omega_{pe}$ is the Debye length, $c_s = \lambda_D \omega_{pi} = \sqrt{k_B T_e/m_i}$ is the sound speed, and $H=\hbar\omega_{pe}/(k_BT_e)$ is a dimensionless parameter that measures the importance of quantum effects.

\paragraph{Fluid theory}
The relevant 1D fluid equations for the ions and the electrons read as follows
\begin{eqnarray}
\partial_t n_{i,e} &+& \partial_x (n_{i,e} u_{i,e} ) = 0,  \label{eq:continuity_ie}\\
m_e (\partial_t u_{e} &+& u_e \partial_x u_{e} ) = e \partial_x V_H - {{\partial_x P_e} \over  n_e}  + \zeta\, {\hbar^2 \over{2 m_e}} \partial_x\left(\frac{\partial_x^2 \sqrt{n_e}}{\sqrt{n_e}} \right), \label{eq:euler_elec} \\
m_i (\partial_t u_{i} &+& u_i \partial_x u_i) = -e \partial_x V_H,   \label{eq:euler_ion} \\
\partial_x^2 V_H &=& - {e \over \varepsilon_0} (n_i - n_e) , \\
P_e &=& n_0 k_B T_e \left( {n_e \over n_0}\right)^\gamma ,
\end{eqnarray}
where we already neglected thermal and quantum effects for the ions. Further, we also neglect the electron inertia ($m_e \ll m_i$), so that the left-hand side of Eq. \eqref{eq:euler_elec} vanishes.

We write all quantities as the sum of a homogeneous equilibrium term plus a small fluctuation: $n_{e,i} = n_0 + n_{e,i\,1}(x,t)$, and similarly for $u_{e,i} $, $V_H$, and $P_e$. At equilibrium, $u_0=V_{H, 0}=0$ and $P_{e1} = \gamma n_0 k_B T_e$. Then, we write all fluctuations as plane waves, as was done in Eq. \eqref{eq:planewaves}.
Looking for normal modes, we obtain after some algebra:
\be
\omega_{\rm fluid}^2 = \frac{c_s^2 k^2 + \zeta{H^2 \over 4} k^4 \lambda_D^4 \omega_{pi}^2}{1+\gamma k^2 \lambda_D^2 + \zeta{H^2 \over 4} k^4 \lambda_D^4 \omega_{pi}^2}
\label{eq:disp_relation_acoustic_fluid}
\ee
where $c_s = \sqrt{\gamma k_B T_e/m_i}$ is the sound speed.

In order to compare the kinetic and fluid results, we expand Eqs. \eqref{eq:disp_relation_acoustic_kin} and \eqref{eq:disp_relation_acoustic_fluid} in powers of the wave number $k$ and retain terms up to $k^4$. We obtain:
\begin{eqnarray}
\frac{\omega_{\rm kin}^2}{c_s^2 k^2 } &= 1 - \left(1-\frac{H^2}{12}\right) k^2 \lambda_D^2 + \dots \\
\frac{\omega_{\rm fluid}^2}{c_s^2 k^2 } &= 1 - \left(\gamma-\zeta \frac{H^2}{4}\right) k^2 \lambda_D^2 + \dots ,
\end{eqnarray}
which coincide for $\gamma=1$ and $\zeta =1/3$. The exponent $\gamma=1$ can be explained by the ions evolving in the thermal bath of the electron population, which is at fixed temperature $T_e$. Hence, the ion fluid follows an isothermal transformation, which is indeed characterized by $\gamma=1$.

\subsection{Discussion}
From the above examples, it appears there is no universal closure relation that is valid for all regimes. Restricting ourselves to the polytropic form \eqref{eq:polytropic} for the classical pressure and the expression \eqref{pression Q n_iso} -- or equivalently the Bohm potential \eqref{bohm potential} -- for the quantum pressure, yields a family of closure relations that depend on the two parameters $\gamma$ and $\zeta$. In the linear regime, the values of these two parameters depend on the frequency and the wavelength of the normal modes under consideration. Here, we have seen that the extreme cases $\omega/k \gg v_{th,e}$ (Langmuir waves) and $\omega/k \ll v_{th,e}$ (ion acoustic waves) yield different values for these two parameters, as was already noticed in \cite{Haas2015}.

The closure relations adopted here are all local, ie. they depend on the value of the electron density at a certain spatial point.
Moldabekov et al. \cite{Moldabekov2018} provided a general framework for such local closure relations in the various regimes of short or long wavelength, and small or large frequency. In addition, they generalized the Bohm potential to include non-local effects. Very recently, the same authors introduced the concept of many-body Bohm potential and showed how it can be used to improve the accuracy of the quantum fluid description \cite{Moldabekov2021}.

{\gm
In the sections above, we discussed the closure relations for the pressure and Bohm potential for a 3D electron gas (although the problem was further reduced to 1D, the assumptions for the closure assume a 3D gas).
In connection to nanophysics, there exist materials where the electron dynamics effectively occurs in lower dimensions. Examples include 2D electron gases occurring in semiconductor heterostructures \cite{Khan1992} or graphene layers \cite{Berger2004}, and 1D electron gases in nanowires and carbon nanotubes \cite{Tans1997}.
For these low-dimensional configurations, the closure relations for the quantum pressure (Bohm potential)  are fundamentally different from the 3D case, and the resulting dispersion relations are different too. These issues were discussed in a recent work \cite{Moldabekov2017}.
}

We conclude this section by noting the similarity between the quantum fluid approach and the time-dependent {\it orbital-free density functional theory} (OF-DFT) \cite{Witt2018OFDFT}.
The latter is a description of the electron gas that is based uniquely on the electron density $n(\bm{r})$ (like all versions of DFT), without requiring the calculation of pseudo-orbitals as is done in the more standard Kohn-Sham approach (for a brief discussion of DFT and its relationship to quantum plasmas, see \cite{Manfredi2019}).
The theorems of DFT ensure that a complete description of the electron gas can be obtained through the electron density alone, provided one is able to write down the exact energy functional
\be
E[n] = T[n]+ E_{ext}[n] + E_H[n]+ E_{XC}[n]  ,
\label{enfunctional}
\ee
where the various terms on the right-hand side represent, respectively, the kinetic, external, Hartree, and exchange-correlation energies.

Next, we suppose that the kinetic energy functional may be written as $T[n] = T_W[n] + T_{\theta}[n]$, i.e. the sum of a  von Weizs\"acker energy
\be
T_W[n] = \int \sqrt{n}\left({-\hbar^2 \over {2m}}\right)\nabla^2 \sqrt{n}\,d\bm{r} ,
\label{eq:vW}
\ee
and a residual term $T_{\theta}[n]$ (known as the Pauli functional, which will incorporate, among others, terms that pertain to the ``quantum pressure" in the fluid formalism).
Then, the function $\Psi \equiv \sqrt{n}$ obeys a Schr\"odinger-like equation with a density-dependent effective potential
\be
V_{eff}[n] = \frac{\delta}{\delta n} \Big(T_{\theta}[n]+ E_{ext}[n] + E_H[n]+ E_{XC}[n]\Big),
\ee
where $\frac{\delta}{\delta n} $ denotes the functional derivative  \cite{Levy1984,Chan2001}.
By generalizing to the time-dependent case and performing an inverse Madelung transformation, i.e. by writing $\Psi = \sqrt{n}\, e^{iS/\hbar}$ [see Eq. \eqref{madelung decomposition}], one can derive a system of equations for $n$ and $\bm{u} = \nabla S/m$ which is identical to our quantum fluid model. The theorems of DFT and TDDFT then ensure that it exists, at least in principle, an energy functional which renders the density-dependent equations exactly equivalent to the full N-body problem.

All this may seem in contradiction with the above statement that there is no universal closure relation for the fluid equations. However, for the time-dependent version of DFT, the energy functional may depend on the initial condition, hence different evolutions (Langmuir and acoustic waves, for instance) could  require different functionals. Further work on the comparison between quantum fluid theory and time-dependent OF-DFT is needed to clarify those subtle issues.


\section{Application: Electronic breathing modes in metallic nanoshells}
\label{metal_nanoshells}

{\gm Here, we apply the quantum fluid equation derived in section \ref{Fluid models} to a particular class of nano-objects, namely metallic nanoshells, which are hollow structures with approximately spherical geometry, and radius in the range 20--100 nm.\footnote{\gm Parts of this section appeared earlier as a conference proceeding \cite{Manfredi2018SPIE}. They are reproduced here with permission.}
}

\subsection{Model}
As a minimal quantum fluid model for metallic nanoshells \cite{Manfredi2018SPIE}, we consider a spherically-symmetric system where all quantities depend only on the radial coordinate $r$ and the time $t$. The ion lattice is represented by a uniform continuous positive charge density (jellium), whereas the electrons are described by the following set of fluid equations. Atomic units (a.u.) are used in this section; these correspond to setting $4\pi \varepsilon_0=e=\hbar=1$ in the formulas written in SI units. In atomic units and spherical co-ordinates, the continuity and Euler equations read as:
\begin{eqnarray}
&& \frac{\partial n}{\partial t} + {1 \over r^2}\frac{\partial}{\partial r}\left(n u r^2 \right)=0 \,, \label{eq:continuity_shell}\\
&& \frac{\partial u}{\partial t}+u\frac{\partial u}{\partial r}  = \frac{\partial V_H}{\partial r}+\frac{1}{2}\frac{\partial}{\partial r} \left(\frac{\nabla^2_r\sqrt{n}}{\sqrt{n}}\right)-
{1 \over n}\frac{\partial P}{\partial r}
-\frac{\partial V_{XC}}{\partial r}, \label{eq:euler}\
\end{eqnarray}
where $n$ is the electron number density, $u$ is the mean radial velocity, $P$ is the (isotropic) pressure, $V_{XC}$ is the exchange and correlation potential. $V_H$ is the Hartree potential obtained from Poisson's equation:
\(
\nabla^2_r V_H = 4\pi \left(n-n_i \right) \label{eq:poissonfluid},
\)
where $n_i(r)$ is the density of the ion jellium and $\nabla^2_r = {1 \over r^2}\partial_r (r^2 \partial_r)$ is the radial Laplacian operator.

The second term on the right-hand side of Eq. \eqref{eq:euler} is the  Bohm potential, which contains quantum effects to lowest order.
For the exchange potential, we use the standard local density approximation (LDA):
\begin{equation}\label{eq:Vx}
V_X [n]= -\frac{(3\pi^2)^{1/3}}{\pi}\, n^{1/3},
\end{equation}
and for the correlations we employ the functional proposed by
Brey et al.\cite{Brey1990}, which yields the following correlation potential:
\begin{equation}\label{eq:Vc}
V_{C}[n] = -\gamma \ln \left(1+\delta n ^{1/3}\right),
\end{equation}
with $\gamma = 0.03349$ and $\delta = 18.376$ in atomic units.
We further assume the electron temperature to be much lower than the Fermi temperature of the metal, so that the pressure can be approximated by that of a fully degenerate electron  gas, as in Eq. \eqref{pressure classical}.

With the aim of modelling  metallic nanoshells, the ion jellium density $n_i(r)$ is chosen to be constant and equal to $n_0$ inside a spherical shell of internal radius $R_i$ and external radius $R_e$, and zero outside. We further define the nanoshell mean radius $R=(R_i+R_e)/2$ and the thickness $\Delta=R_e-R_i$. Assuming global charge neutrality, the total number of electrons inside the shell is:
\begin{equation}\label{eq:Nelectrons}
N = n_0\, \mathcal{V} =\frac{R_e^3-R_i^3}{r_s^3},
\end{equation}
where $\mathcal{V}$ is the volume of the nanoshell and $r_s$ is the Wigner-Seitz radius of the metal. Finally, the plasmon frequency can be written as: $\omega_p=\sqrt{3/r_s^3}$.
In the forthcoming sections, we will consider sodium nanoshells with $r_s=4$.

\subsection{Numerical results: ground state}

\label{sec:groundstate}
   \begin{figure} [ht]
   \begin{center}
   \includegraphics[height=4.5cm]{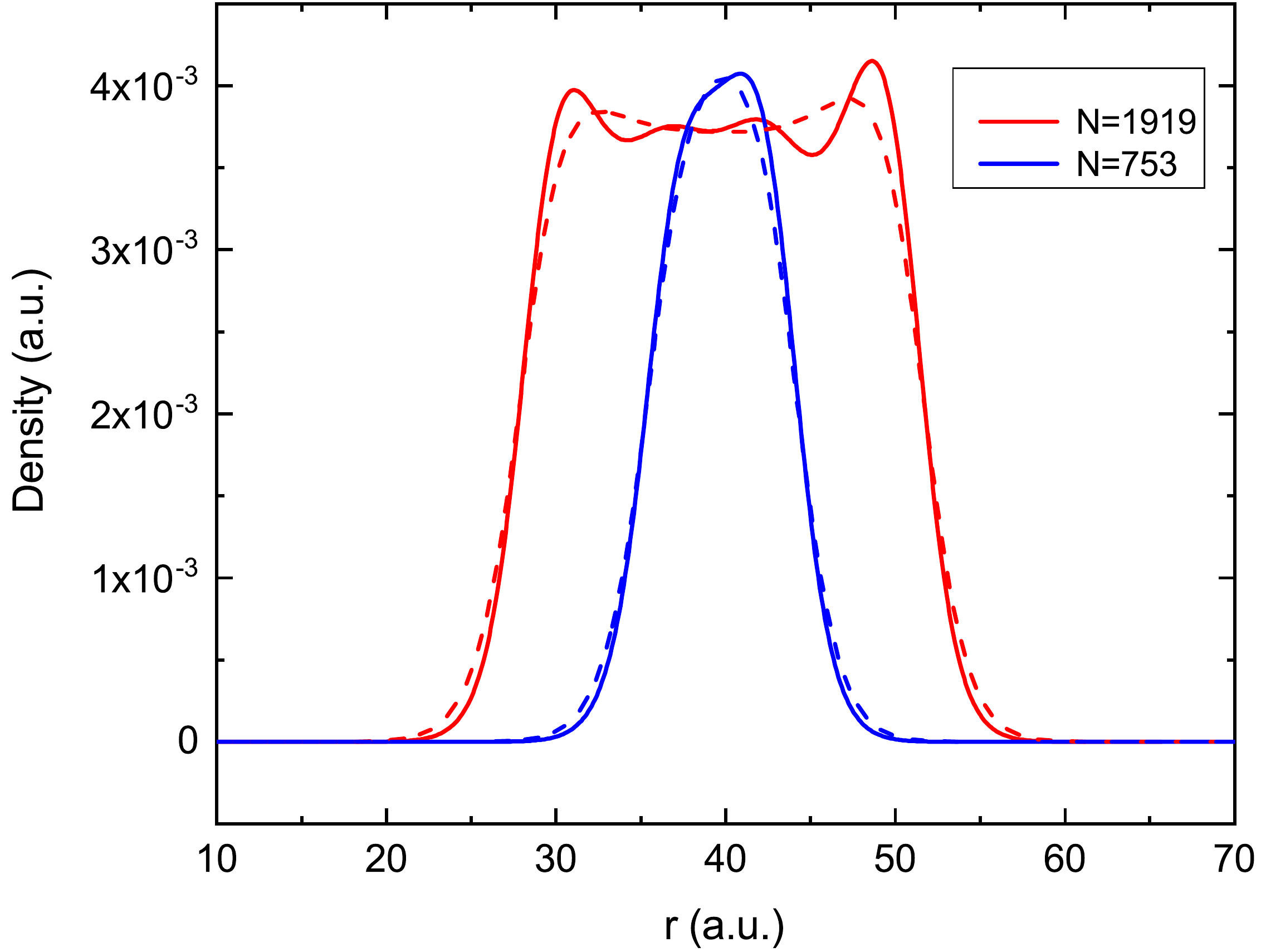}
   \includegraphics[height=4.5cm]{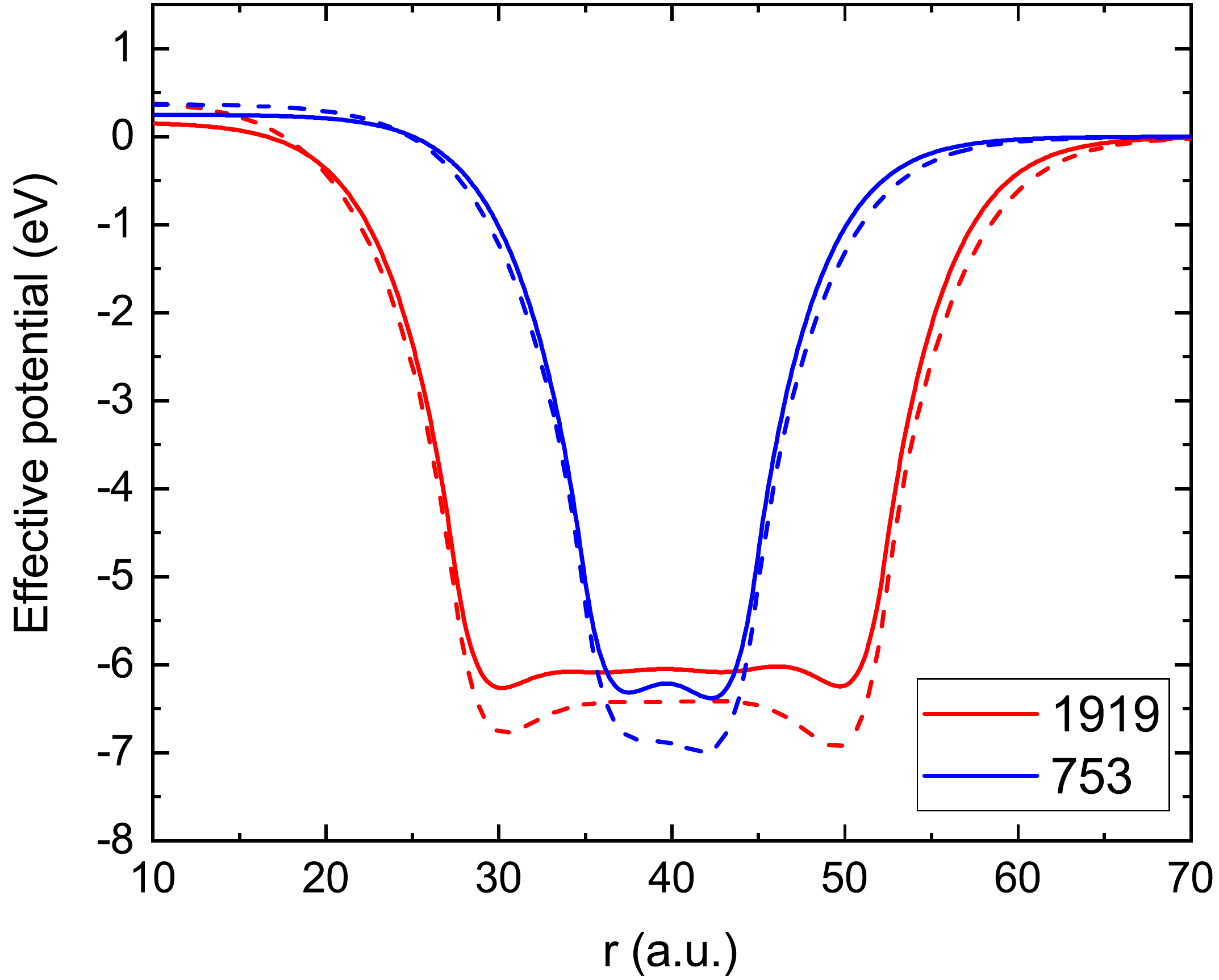}
   \end{center}
   \caption
   { \label{fig:gshydro}
Ground state of two typical Na nanoshells  containing $N=1919$ and $N=753$ electrons. Left panel: Electron densities computed using a DFT code (solid lines) and the fluid  approach (dashed lines). Right panel: Effective potentials for the same cases. Reprinted with permission from \cite{Manfredi2018SPIE}.
}
   \end{figure}

Before considering the dynamical response of the system, we need to compute its ground state. This can be obtained as a stationary solution of Eqs. \eqref{eq:continuity_shell}-\eqref{eq:euler} which is computed numerically using an iterative relaxation procedure \cite{Crouseilles2008}.

The electron density and effective potential profiles for two typical Na nanoshells containing respectively $N=1919$ and $N=753$ electrons are presented in Fig. \ref{fig:gshydro}. Both nanoshells have a central radius $R = 40\, \rm a.u.$, but different thicknesses ($\Delta=10$ and $25\, \rm a.u.$).
On the same figure, we also show the density profiles computed using a standard DFT code. The agreement is very good, and even more so considering that the fluid results (labeled QHD) require no more than a few minutes runtime on a standard desktop computer. In particular, the nonlocal spillout effect (electron density extending beyond the steplike ion density profile) is well described by the QHD method, even for the smaller structure ($N=753$), where the spillout is very prominent and the electron density is nowhere flat.
For the larger nanoshell ($N=1919$), the two overdensities at the internal and external radii in the QHD density profile are a lower-order quantum effect, a remnant of the well-known Friedel oscillations visible in the corresponding DFT profile.

\subsection{Numerical results: dynamics}

Having computed the ground state of the electron system, we need to perturb it slightly in order to induce some dynamical behavior. As we are interested in plasmonic breathing modes, the perturbation will also be spherically symmetric.
For the excitation, we use an instantaneous Coulomb potential applied at the initial time:
\(
V_{ext}(r,t) = \frac{z}{r}\, \tau\,\delta(t),
\)
where $\delta$ is the Dirac delta function, $z$ is a fictitious charge quantifying the magnitude of the perturbation, and $\tau$ is the duration of the pulse.

\begin{figure} [ht]
   \begin{center}
   \includegraphics[height=4.5cm]{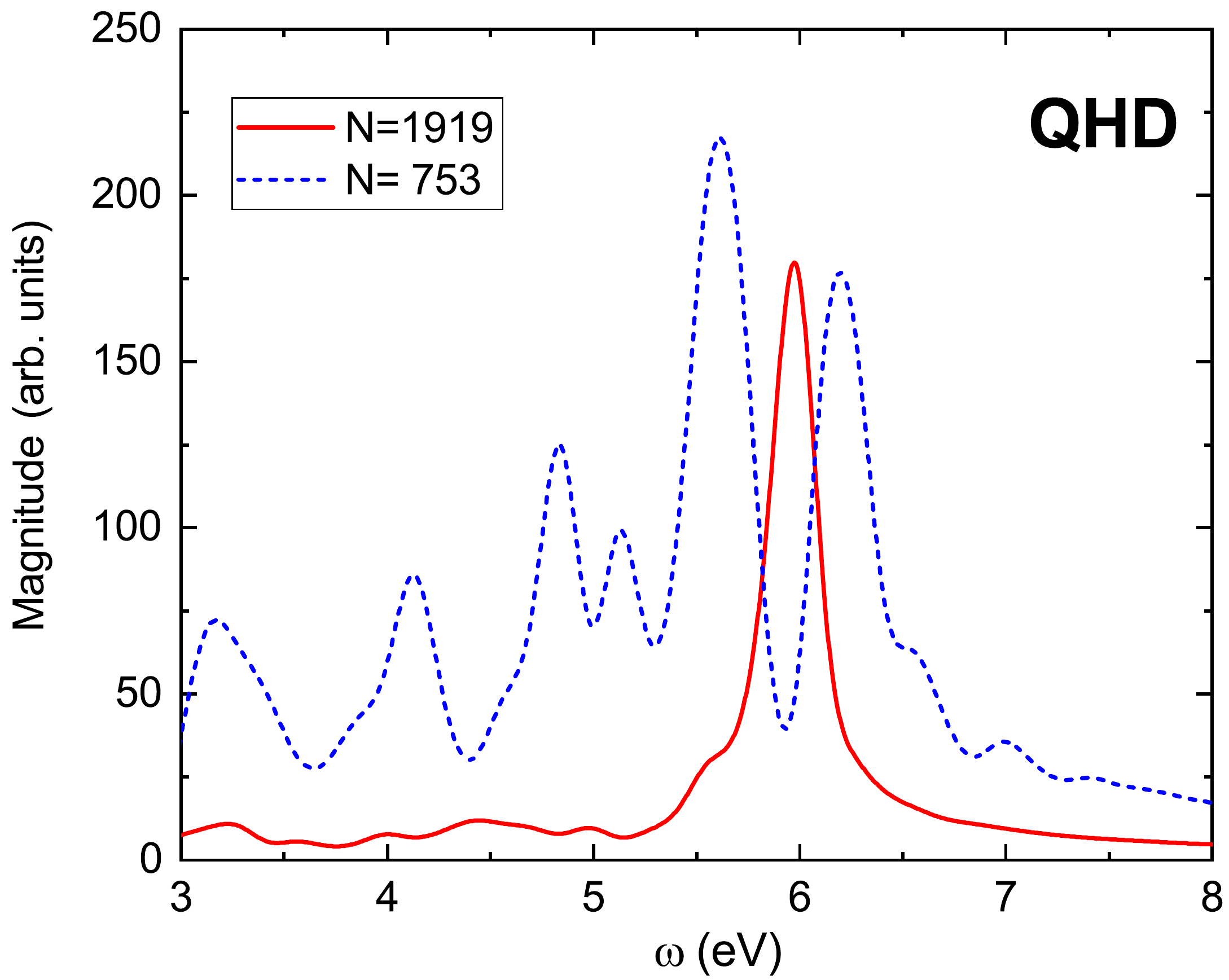}
 \includegraphics[height=4.5cm]{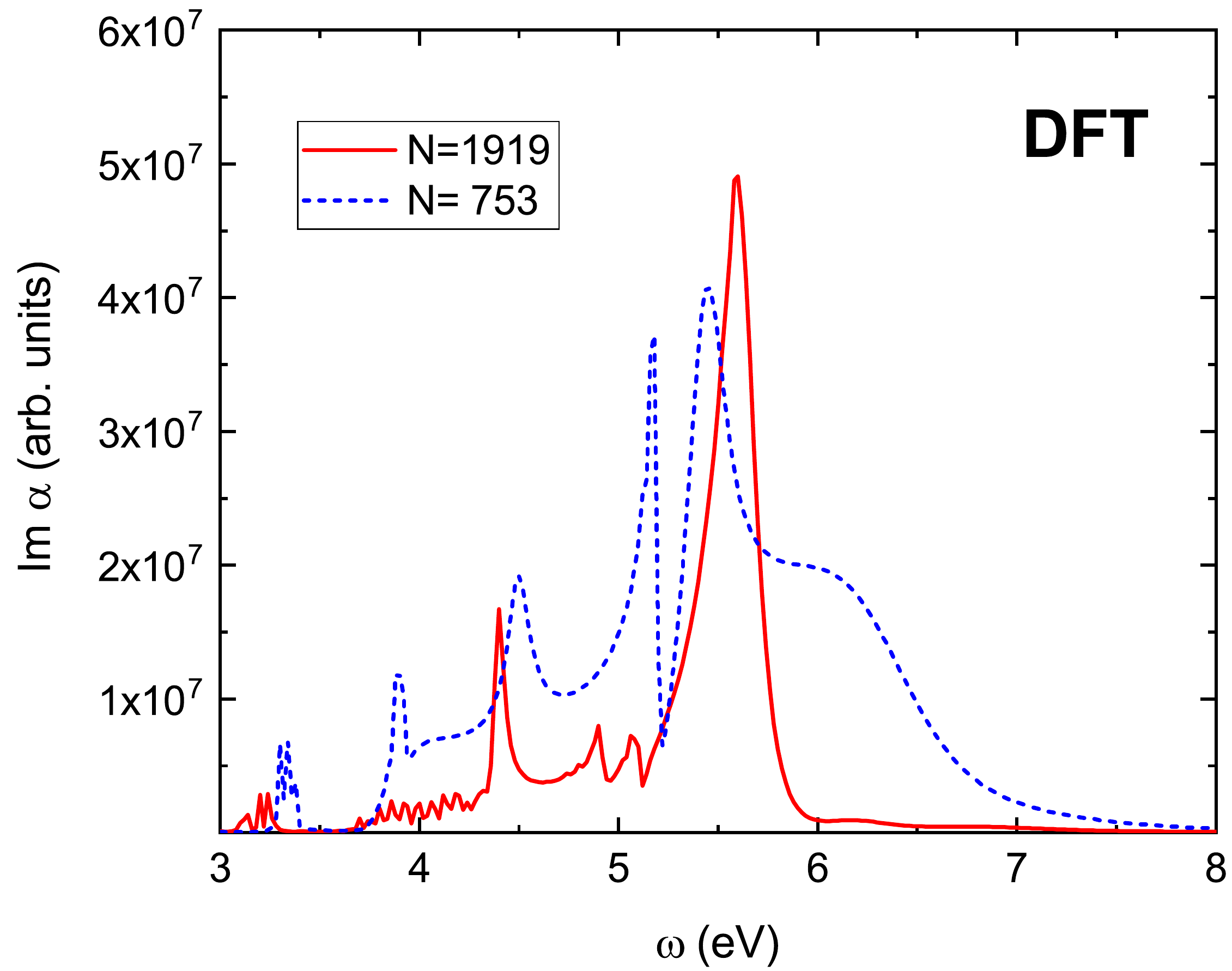}
   \end{center}
   \caption{ \label{fig:spectra}
Linear-response frequency spectrum  for two typical Na nanoshells containing $N=1919$ (red solid lines) and $N=753$ (blue dotted lines) electrons. Left frame: fluid (QHD) results; Right frame: TDDFT results. Reprinted with permission from \cite{Manfredi2018SPIE}.
}
   \end{figure}

In order to analyze the linear response of the system, we study the evolution of the mean radius of the electron cloud, defined as:
\(
\langle r \rangle= {1\over N} \int_0^{\infty} r n(r,t) \,4\pi r^2 dr.
\)
The Fourier transform of $\langle r \rangle$ in the frequency domain is shown in Fig. \ref{fig:spectra}, left frame. For the larger nanoshell ($N=1919$), the frequency spectrum shows a sharp peak near the plasmon frequency $\omega_p =\sqrt{3/r_s^3}= 5.89$~eV. This behavior is compatible with the computed ground-state density profile (Fig. \ref{fig:gshydro}), which displays a region of almost constant density in between the inner and outer radii.
In contrast, the spectrum of the smaller nanoshell ($N=753$) is much more fragmented and actually displays {\em two} principal peaks around the plasmon frequency, at about 5.6~eV and 6.2~eV, plus a number of smaller peaks at lower energies.

\label{sec:groundstate}

Still on Fig. \ref{fig:spectra} (right panel), we show the monopolar polarizability $\alpha$ computed with a linear-response TDDFT code \cite{Maurat2009}, using the same parameters and exchange-correlation functionals as for the fluid (QHD) simulations. The results compare rather favorably with the fluid ones. For $N=1919$, one dominant peak is observed at 5.6~eV, i.e. slightly redshifted compared to the pure plasmon frequency. This redshift can be understood in terms of dissipative phenomena (such as Landau damping, i.e. the coupling of the plasmon mode to single-particle modes) that are not included in the fluid description.
More interestingly, the spectrum of the smaller nanoshell ($N=753$) reveals the same two-peak structure also observed in the fluid simulations, now with frequencies $\approx 5.16 \,\rm eV$ and 5.45~eV, again slightly redshifted compared to corresponding peaks in the fluid spectrum.
The more complex spectrum observed for $N=753$ is probably due to the shape of the ground state electron density, which looks more like a bell curve  with no flat region inside the lattice jellium (see Fig. \ref{fig:gshydro}).


\section{Fluid models with spin effects} \label{Fluid model with spin effects}

The electrons are elementary fermions that carry not only an electric charge equal to $-e$, but also a magnetic moment (spin) equal to $\hbar/2$. The effect of the spin appears whenever the electron is immersed in a magnetic field. Even in the absence of an external  field, an electron orbiting in the Coulomb potential of a positively-charged nucleus will feel the effect of a magnetic field in its own frame of reference (this  ``spin-orbit coupling" will be discussed in Sec. \ref{sec:spinorbit}).
Spin effects play an important role in many modern applications aimed at the storage and transfer of information, which go under the name of spintronics \cite{Hirohata2020}.
In plasma physics, polarized electron beams can now be created and precisely manipulated in the laboratory \cite{Wu2019,Wu2020,Nie2021,Crouseilles2021}.

{\gm Some of the material presented in this section is taken from the PhD thesis of one of the co-authors \cite{Hurst2017phd}.}

\subsection{General spin fluid equations}

Fluid models that include spin effects are derived, as usual, from the corresponding kinetic (Wigner or Vlasov) equation by taking velocity moments.
For spin-1/2 fermions, the Wigner function is no longer a scalar, but rather a $2 \times 2$  matrix \cite{Arnold1989}. The fully quantum evolution equation of such matrix Wigner function is extremely complicated and hence of limited practical use.
In the semiclassical limit, this equation gives rise to a matrix spin-Vlasov equation, which treats the electron motion in a classical fashion while preserving the intrinsically quantum character of the spin degrees of freedoms \cite{Hurst2014,Hurst2017}:
\begin{align}
&\frac{\partial f_{0}}{\partial t}
+
\bm{v} \cdot \bm{\nabla} f_{0}  - \frac{e}{m} \left( \bm{E} + \bm{v} \times \bm{B} \right) \cdot \bm{\nabla_{v}} f_{0} + \frac{\mu_B }{2mc^{2}} \left( \bm{E} \times \bm{\nabla} \right)_{i} f_{i} \nonumber \\
&~
- \frac{\mu_B}{m} \bm{\nabla} \left[ B_{i} - \frac{1}{2c^{2}}\left( \bm{v} \times \bm{E} \right)_{i} \right] \cdot \bm{\nabla_{v}} f_{i}  - \frac{\mu_B e}{2m^{2}c^{2}} \left[ \bm{E} \times \left( \bm{B} \times \bm{\nabla_{v}} \right) \right]_{i} f_{i} = 0.
\label{vlasov equation f0 avec spin orbit} \\ \nonumber \\
&
\frac{\partial f_{i}}{\partial t}
+
\bm{v} \cdot \bm{\nabla} f_{i}  - \frac{e}{m} \left( \bm{E} + \bm{v} \times \bm{B} \right) \cdot \bm{\nabla_{v}} f_{i} + \frac{\mu_B }{2mc^{2}} \left( \bm{E} \times \bm{\nabla} \right)_{i} f_{0} \nonumber \\
&~
- \frac{\mu_B}{m} \bm{\nabla} \left[ B_{i} - \frac{1}{2c^{2}}\left( \bm{v} \times \bm{E} \right)_{i} \right] \cdot \bm{\nabla_{v}} f_{0}  - \frac{\mu_B e}{2m^{2}c^{2}} \left[ \bm{E} \times \left( \bm{B} \times \bm{\nabla_{v}} \right) \right]_{i} f_{0}  \nonumber \\
&~
- \frac{2\mu_B}{\hbar} \left\{ \left[ \bm{B} - \frac{1}{2 c^{2}} \left(\bm{v} \times \bm{E} \right) \right] \times \bm{f} \right\}_{i} = 0 ,
\label{vlasov equation f avec spin orbit}
\end{align}
where $( f_0, \bm{f})$ are the four components of the $2 \times 2$ matrix Wigner function, and $\bm{f} = (f_x,f_y,f_z)$. The factor $\hbar$ is hidden in the definition of the Bohr magneton  $\mu_B = e \hbar /(2m)$. The quantum corrections in Eqs. \eqref{vlasov equation f0 avec spin orbit}- \eqref{vlasov equation f avec spin orbit} (terms preceded by $\mu_B$) couple the orbital ($f_0$) and the spin ($f_i$) components of the Wigner function through the Zeeman effect and the spin-orbit interactions. The latter are given by those terms preceded by $1/c^2$, signalling that the spin-orbit coupling is a relativistic effect. There are no quantum corrections to the orbital electron dynamics because they appear only at the second order in $\hbar$.

Starting from the spin-Vlasov Eqs. \eqref{vlasov equation f0 avec spin orbit} - \eqref{vlasov equation f avec spin orbit}, we derive the fluid equations  by taking velocity moments of the phase-space distribution functions.
At first, we will only include the Zeeman interaction. To obtain the fluid closure, we will employ a general procedure based on the maximization of entropy (see section \ref{sec:maxentropy}). Fluid models with spin-orbit effects will be discussed further in section \ref{sec:spinorbit}.

Going back to the Vlasov equations \eqref{vlasov equation f0 avec spin orbit}-\eqref{vlasov equation f avec spin orbit}, we note that the scalar and vector distribution functions $f_0(\bm{r},\bm{v}, t)$ and $\bm f(\bm{r},\bm{v}, t)$ represent, respectively, the probability density for a particle to be at a point $(\bm{r},\bm{v})$ in the phase space and the probability density for that particle to have a spin directed along the direction of $\bm f$.
Hence $f_0$ incorporates all the information unrelated to the spin, while $\bm f$ incorporates all information about the spin of the electron.

With these definitions, the particle density $n $ and the spin polarization $\bm{S} $ of the electron gas are easily expressed as moments of the distribution functions  $f_0$ and $\bm f$:
\begin{eqnarray}
n(\bm{r},t)
&=&
\int f_{0} (\bm{r},\bm{v}, t) d\bm{v}, \label{def n} \\
\bm S (\bm{r},t)
&=&
\frac{\hbar}{2}
\int \bm f (\bm{r},\bm{v}, t)  d\bm{v}.\label{def S}
\end{eqnarray}
We further define the quantities
\begin{eqnarray}
\bm{u} &=& \frac{1}{n}\int \bm{v} f_{0}d\bm{v},\label{def u} \\
J^{S}_{i\alpha}&=& \frac{\hbar}{2} \int v_{i} f_{\alpha}d\bm{v},\label{def J}\\
P_{ij}&=& m\int w_{i} w_{j} f_{0}d\bm{v},\label{def P}\\
\Pi_{ij\alpha} &=& \frac{\hbar}{2} m  \int v_{i} v_{j}  f_{\alpha}  d\bm{v},\label{def Pi}\\
Q_{ijk} &=& m \int w_{i} w_{j} w_{k} f_{0} d\bm{v},\label{def Q}
\end{eqnarray}
where we separated the mean fluid velocity $\bm u$ from the velocity fluctuations $\bm{w} \equiv \bm{v} - \bm{u}$.
Here, $P_{ij}$ and $Q_{ijk} $ represent the pressure and the generalized energy flux tensors. They coincide with the corresponding definitions for spinless fluids with probability distribution function $f_0$.
The spin-velocity tensor $J_{i \alpha}^S$ represents  the mean fluid velocity along the $i$-th direction of the $\alpha$-th spin polarization vector, while $\Pi_{ij\alpha}$ represents the corresponding spin-pressure tensor\footnote{Strictly speaking a pressure tensor should be defined in terms of the velocity fluctuations $w_i w_j$, but this would unduly complicate the notation. Thus, we stick to the above definition of $\Pi_{ij\alpha}$ while still using the term ``pressure" for this quantity.}.
The evolution equations for the above fluid quantities are  obtained by taking velocity moments of Eqs. \eqref{vlasov equation f0 avec spin orbit} - \eqref{vlasov equation f avec spin orbit}:
\begin{align}
&\frac{\partial n}{\partial t} + \bm{\nabla} \cdot \left(n\bm{u}\right)   = 0,  \label{f0_continuity} \\
&\frac{\partial S_{\alpha}}{\partial t} + \partial_{i} J^{S}_{i \alpha} + \frac{e}{ m } \left( \bm{S} \times \bm{B} \right) _{\alpha} =  0, \\
\label{falpha_continuity}
&\frac{\partial u_{i} }{\partial t} + u_{j} (\partial_{j} u_{i}) + \frac{1}{nm} \partial_{j} P_{ij} + \frac{e}{m} \left[E_{i} + \left(\bm{u} \times \bm{B} \right)_{i}  \right]+\frac{e}{nm^{2}}  S_{\alpha} \left(\partial_{i} B_{\alpha} \right)
= 0,  \\
&\frac{\partial J^{S}_{i \alpha}}{\partial t} + \partial_{j} \Pi_{ij\alpha} + \frac{e E_{i}}{ m }S_{\alpha} + \frac{e}{ m } \epsilon_{jki} B_{k} J^{S}_{j\alpha} + \frac{e}{ m } \epsilon_{jk\alpha} B_{k} J^{S}_{ij} + \frac{\mu_B \hbar}{2m}  \left(\partial_{i} B_{\alpha} \right) n =  0,\label{evol Js}\\
&\frac{\partial P_{ij}}{\partial t}  + u_{k}  \partial_{k} P_{ij}  + P_{jk} \partial_{k} u_{i} +  P_{ik} \partial_{k} u_{j}  +  P_{ij} \partial_{k}  u_{k}    + \partial_{k} Q_{ijk} + \frac{e}{m} \big{[} \epsilon_{lki} B_{k} P_{jl}  \nonumber \\
&\hspace{1cm} + \epsilon_{lkj} B_{k} P_{il} \big{]}   + \frac{e}{m^{2}} \sum_{\alpha} \left[   \partial_{i} B_{\alpha}  \left( J^{S}_{j \alpha} - S_{\alpha} u_{j} \right) +   \partial_{j} B_{\alpha}  \left(J^{S}_{i \alpha} - S_{\alpha} u_{i} \right) \right] = 0.
\label{eq_presure}
\end{align}

A different set of fluid equations for spin-1/2 particles was derived by Brodin and Marklund \cite{Brodin2007} using a Madelung transformation on the Pauli wave function.
Another fluid theory was derived by Zamanian et al. \cite{Zamanian2010} from a Vlasov equation that includes the spin as an independent variable in an extended phase space \cite{Zamanian2010nj}. More recently, a relativistic hydrodynamic model was obtained by Asenjo et al. \cite{Asenjo2011} from the Dirac equation. These approaches usually lead to cumbersome equations that are in practice very hard to solve, either analytically or numerically, even in the nonrelativistic limit.

{\gm
Still another approach is based on a generalization of the Bloch equations \cite{Andreev2015}. Instead of adding ``spin-dependent" moments to the usual ones as was done above -- see Eqs. \eqref{def S}, \eqref{def J}, and \eqref{def Pi} -- Andreev \cite{Andreev2015} treats separately the spin-up and spin-down components in the corresponding Pauli equation. Performing a Madelung transformation for each component, one arrives at a set of four fluid equations (two continuity and two Euler equations), which are then closed using an appropriate equation of state for the pressure.
}

Going back to our fluid  model \eqref{f0_continuity}-\eqref{eq_presure}, some further hypotheses are needed to close the set of equations.
We first note that, by definition, the following equation is always satisfied:
$\int w_{i} f_{0} d\bm{v} = 0$.
The same is not true, however, for the expression obtained by replacing $f_0$ with $f_{\alpha}$ in the preceding integral. If we assume that such a quantity indeed vanishes, i.e.
$\int w_{i} f_{\alpha} d\bm{v} =0$,
we immediately obtain that
\begin{align}
J^{S}_{i\alpha} = u_{i} S_{\alpha}. \label{int clos}
\end{align}
Physically, this means that the spin  is simply transported along the average fluid velocity.
This is of course an approximation that amounts to neglecting some spin-velocity correlations \cite{Zamanian2010}.

With this assumption,  Eq. \eqref{evol Js} is no longer necessary and the system of fluid equations reduces to
\begin{align}
& \frac{\partial n}{\partial t} + \bm{\nabla} \cdot \left(\bm{u} n\right)    =     0, \label{density_first_closure}\\
&\frac{\partial S_{\alpha}}{\partial t} + \partial_{i} \left(u_{i} S_{\alpha}\right) + \frac{e}{ m } \left( \bm{S} \times \bm{B} \right) _{\alpha}   = 0, \label{spin_first_closure} \\
&\frac{\partial u_{i} }{\partial t} + u_{j} (\partial_{j} u_{i}) + \frac{1}{nm} \partial_{j} P_{ij} + \frac{e}{m} \left[E_{i} + \left(\bm{u} \times \bm{B} \right)_{i}  \right]+\frac{e}{nm^{2}}  S_{\alpha} \left(\partial_{i} B_{\alpha} \right)   =  0, \label{velocity_first_closure} \\
&\frac{\partial P_{ij}}{\partial t}  + u_{k}  \partial_{k} P_{ij}  + P_{jk} \partial_{k} u_{i} +  P_{ik} \partial_{k} u_{j}  +  P_{ij} \partial_{k}  u_{k}    + \partial_{k} Q_{ijk} \nonumber \\
&\hspace{2cm} + \frac{e}{m} \big{[} \epsilon_{lki} B_{k} P_{jl}   + \epsilon_{lkj} B_{k} P_{il} \big{]}  = 0.
\label{eq_presure2}
\end{align}
Interestingly, in Eq. \eqref{spin_first_closure} the spin polarization is now transported by the fluid velocity $\mathbf{u}$, as in the model of Zamanian et al. \cite{Zamanian2010}.
In order to complete the closure procedure, one can proceed in the same way as is usually done for spinless fluids, see section \ref{Fluid models}, for instance by assuming that the pressure is isotropic and replacing Eq. \eqref{eq_presure2} with the polytropic expression \eqref{eq:polytropic}.

\subsection{Fluid closure: Maximum entropy principle} \label{sec:maxentropy}

The maximum entropy principle (MEP) is a well-developed theory that has been successfully applied to various areas of gas, fluid, and solid-state physics \cite{Ali2012,Trovato2010,Romano2001,Anile1995}. The underlying assumption of the MEP is that, at equilibrium, the probability distribution function is given by the most probable microscopic distribution  (i.e., the one that maximizes the entropy) compatible with some macroscopic constraints. The constraints are generally given by the various velocity moments, i.e., the local density, mean velocity, and temperature.

Here, we shall follow the derivation detailed in  \cite{Hurst2014}.
In order to illustrate the application of the MEP theory to a spin system, we write the  Hamiltonian as follows
\begin{equation}
  \mathcal{H}= m \frac{|\bm{v}|^2}{2}   + V(\bm{r})  \sigma_0 +\mu _{B}  \bm{B} \cdot \bm{\sigma},
\end{equation}
where $\sigma_0$ is the $2 \times 2$ identity matrix and  $\bm{\sigma}$ is the vector of the Pauli matrices
\be
\sigma_x = \begin{pmatrix}
0 & 1 \\ 1&0 \end{pmatrix}, \,\,\,
\sigma_y = \begin{pmatrix}
0 & -i \\ i&0 \end{pmatrix}, \,\,\, \label{eq:paulimatrices}
\sigma_z = \begin{pmatrix}
1 & 0 \\ 0&-1\end{pmatrix} .
\ee

The relevant entropy  density is
\begin{equation}
 s(\mathcal{F})=\left\{
 \begin{array}{ll}
               k_B ~Tr \left\{ \mathcal{F} \log \mathcal{F} -\mathcal{F} \right\} &  ~~\textrm{(M--B)}\\[2mm]
               k_B~ Tr \left\{ \mathcal{F} \log \mathcal{F}+ (1-\mathcal{F}) \log (1-\mathcal{F}) \right\} & ~~\textrm{(F--D),}
              \end{array}\right\} ,
\label{entropy distribution function}
\end{equation}
for Maxwell-Boltzmann (M-B) and Fermi-Dirac (F-D) statistics, respectively. Here,
$\mathcal{F}$ is the $2 \times 2$ matrix Wigner function
\begin{equation}
\mathcal{F}= \frac{1}{2}\sigma _0 f_0 + \frac{1}{2}\bm{f} \cdot \bm{\sigma},
\label{change basis wigner function}
\end{equation}
The detailed calculations for three- and four-moment closures  are given in \cite{Hurst2014}.
Here, we consider the simplest case where only three fluid moments (density $n$, mean velocity $\bm{u}$, and spin polarization $ \bm{S}$) are kept,

For the M-B statistics, the pressure and the spin current at equilibrium turn out to be
\begin{eqnarray}
P_{ij} &=& m {\rm Tr} \left( \int v_{i} v_{j} \mathcal{F}^{ eq } d\bm{v}\right) - m n u^{2}  =   n k_B T \delta_{ij} \\
J^{S}_{i \alpha} &=& S_{\alpha} u_{i}.
\end{eqnarray}
Thus, considering three fluid moments and M-B statistics, leads to the standard expression for the isotropic pressure of an ideal gas, together with
the ``intuitive" closure condition \eqref{int clos} for the spin current tensor.

We repeated the above procedure for the F-D statistics and, as in the case of M-B, we recover the closure: $J^{S}_{i\alpha} = u_{i} S_{\alpha}$. For the pressure, we obtain
\begin{eqnarray}
P &=&
\frac{\hbar^{2}}{5m}\frac{\left(6\pi ^{2} \right)^{2/3}}{2^{5/3}} \left[\left( n - \frac{2}{\hbar}|\bm{S}|\right)^{5/3} + \left( n + \frac{2}{\hbar}|\bm{S}|\right)^{5/3} \right].\label{P FD}
\end{eqnarray}
When the spin polarization vanishes, Eq. \eqref{P FD} reduces to the usual expression of the zero-temperature pressure of a spinless Fermi gas \eqref{pressure classical}.
This can be interpreted as the total pressure of a plasma composed by two populations (spin-up and  spin-down electrons). Due to the Zeeman splitting, the density of the particles whose spin is parallel to the magnetic field is lower than the energy of the particles whose spin is antiparallel. Equation \eqref{P FD} shows that the two populations provide a separate contribution to the total fluid pressure.

\subsection{Linear dispersion relation} \label{sec:dispersion_spin}

We want to compute the effect of the spin dynamics on the linear dispersion relation for Langmuir waves. To describe the electron dynamics, we use the following fluid model:
\begin{align}
&\frac{\partial n}{\partial t} + \bm{\nabla} \cdot \left( n \bm{u}\right) = 0,
\label{Eq:fluid_n} \\
&\frac{\partial S_\alpha}{\partial t} + \partial_i \left( u_i S_\alpha \right) + \frac{e}{m} \left( \bm{S} \times \bm{B}\right)_\alpha = 0,
\label{Eq:fluid_S}\\
&\frac{\partial u_i}{\partial t} + u_j \left( \partial_j u_i \right) + \frac{\partial_i P}{n m}  + \frac{1}{m}\partial_i\left( -eV_H +V_{xc}\right) + \frac{e}{n m^2}S_\alpha \partial_i \left( \bm{B}+ \bm{B_{xc}}\right)_\alpha,
\label{Eq:fluid_u}
\end{align}
where we have used the closure relations: $J^S_{i \alpha} = u_i S^\alpha$ for the spin current, and $P = n_0 k_B T (n/n_0)^\gamma$ for the isotropic pressure. The Hartree potential is a solution of Poisson's equation: $\bm{\nabla}^2 V_H = \frac{e}{\epsilon_{0}} ( n-n_0)$, where $n_0$ is the homogeneous ion density.

The above equations are essentially identical to Eqs. \eqref{density_first_closure}-\eqref{velocity_first_closure}, with some extra terms:
\begin{enumerate}
\item
The magnetic field $\bm{B}(\bm{r},t) = \bm{\nabla} \times \bm{A}(\bm{r},t)$ is a solution of the quasi-static  Maxwell-Amp{\`e}re equation
\begin{align}
\bm{\nabla}^2 \bm{A}(\bm{r},t) = -\frac{e}{m}\mu_0 \bm{\nabla} \times \bm{S}(\bm{r},t).
\label{Eq:Maxwell_ampere_equation}
\end{align}
Note that we did not consider the self-consistent magnetic field created by charge currents and neglected the Lorentz force in the momentum equation \eqref{Eq:fluid_u}. The reason is that we  focus on plasma waves propagating in a well-defined longitudinal axis, with no coupling to the other transverse directions. Hence, the only contribution to the magnetic field arises from the spin polarization $\bm{S}$ and accounts for magnetic dipole interactions.
\item
We include  the exchange and correlation fields  $V_{xc}[n(\bm{r},t),\bm{m}(\bm{r},t)]$ and $\bm{B}_{xc}[n(\bm{r},t),\bm{m}(\bm{r},t)]$.
Here, $\bm{m} = 2\bm{S}/\hbar$ is the magnetization vector, which has the same dimensions as the electron density and can be used to compute the electron polarization rate: $\eta = |\bm{m}|/n$. Exchange and correlation potentials can be taken from the vast literature on DFT (see, for instance \cite{Maurat2009} \cite{Gunnarsson1976}) and enable us to go beyond the mean-field approximation, where the only electron-electron interactions are those modeled by the self-consistent Hartree potential. For the present derivation, it is not necessary to provide an explicit form for the exchange and correlation fields.
\end{enumerate}

To obtain the dispersion relation of plasma oscillations, we study the evolution of a small deviations from the equilibrium. The equilibrium state is given by a homogeneous electron gas with density $n_0$  and vanishing mean velocity ($u_0=0$), which is initially spin-polarized along the $z$ direction ($\bm{S} = S_0 \hat{e}_z$). Assuming that the perturbations are plane waves propagating in the $x$ direction, we can express the fluid quantities as follows:
\begin{align}
n(\bm{r},t) &= n_0 + \delta n \exp \left[ i \left( k x - \omega t\right) \right], \label{Eq:delta_n} \\
\bm{S}(\bm{r},t) &= \bm{S}_0 +  \bm{\delta S} \exp \left[ i \left( k x - \omega t\right) \right], \label{Eq:delta_S}  \\
 \bm{u}(\bm{r},t)  &= \bm{\delta u}\exp \left[ i \left( k x - \omega t\right) \right], \label{Eq:delta_u}
\end{align}
where the subscript $0$ denotes  quantities at equilibrium and the $\delta$ quantities are small perturbations.  We write the self-consistent and the exchange-correlation fields in the same way as
\begin{align}
\phi(\bm{r},t) &= \delta \phi \exp \left[ i \left( k x - \omega t\right) \right], \label{Eq:delta_phi}  \\
\bm{B}(\bm{r},t) &=   \bm{\delta B} \exp \left[ i \left( k x - \omega t\right) \right], \label{Eq:delta_B} \\
V_{xc}[n,\bm{m}]  &= V_{xc}[n_0,\bm{m_0}]+ \left(
\delta n \left.\frac{ \partial V_{xc}}{\partial n}\right\vert_{\underset{\bm{m_0}}{n_0}}
+
\frac{2}{\hbar} \bm{\delta S} \cdot \left.\frac{ \partial V_{xc}}{\partial \bm{m}}\right\vert_{\underset{\bm{m_0}}{n_0}}
\right)
 e^{i ( k x - \omega t) }, \label{Eq:delta_Vxc}   \\
B_{xc}[n,\bm{m}]  &= B_{xc}[n_0,\bm{m_0}]+ \left(
\delta n \left.\frac{ \partial B_{xc}}{\partial n}\right\vert_{\underset{\bm{m_0}}{n_0}}
+
\frac{2}{\hbar} \bm{\delta S} \cdot \left.\frac{ \partial B_{xc}}{\partial \bm{m}}\right\vert_{\underset{\bm{m_0}}{n_0}}
\right)
 e^{i ( k x - \omega t) }. \label{Eq:delta_Bxc}
\end{align}

Injecting Eqs. \eqref{Eq:delta_n}-\eqref{Eq:delta_Bxc} into the fluid equations \eqref{Eq:fluid_n}-\eqref{Eq:fluid_u}
and retaining only first-order terms yields the following dispersion relation:
\begin{align}
\omega^2 &= \omega_p^2 + k^2 \gamma \frac{k_B T}{m} +\nonumber\\
& k^2 \frac{n_0}{m}
\left(
\frac{ \partial V_{xc}}{\partial n}+ \eta_0 \frac{ \partial V_{xc}}{\partial m_z}+ \eta_0 \mu_B \frac{ \partial B_{xc}}{\partial n} + \eta_0^2 \mu_B \frac{ \partial B_{xc}}{\partial m_z} \right)
-k^2 \frac{\mu_B^2 \mu_0 n_0}{m} \eta_0^2,
\label{eq:dispertion_relation_spin}
\end{align}
where $\eta_0 = |\bm{m}_0|/n_0$ is the equilibrium polarization.
The second term on the right-hand side is the usual thermal correction (Bohm-Gross), the next term represents the exchange-correlation corrections, and the last term derives from the self-consistent magnetic field [Maxwell-Amp\`ere equation \eqref{Eq:Maxwell_ampere_equation}]. Terms preceded by $\eta_0$ are due to the electron polarization

The above dispersion relation \eqref{eq:dispertion_relation_spin} is identical (in the small wave number $k$ regime) to that obtained from the corresponding spin-Vlasov equations in \cite{Manfredi2019} (see section 7.3), provided that the polytropic exponent is taken $\gamma=3$, as was discussed earlier in section \ref{sec:langmuir}.

\subsection{Spin-orbit coupling} \label{sec:spinorbit}
In the above calculations, we constructed fluids models with spin effects by only considering the Zeeman interaction. The same procedure can be carried over to the case where  spin-orbit interactions are included \cite{Hurst2017}. A straightforward integration of Eqs. \eqref{vlasov equation f0 avec spin orbit}-\eqref{vlasov equation f avec spin orbit}, with respect to the velocity variable, leads to the following fluid equations :
\begin{align}
&\frac{\partial n}{\partial t} + \bm{\nabla} \cdot \left(n\bm{\overline{u}}\right)   = 0,  \label{f0_continuity s-o} \\
&\frac{\partial S_{\alpha}}{\partial t} + \partial_{i} \overline{J}^{S}_{i \alpha} + \frac{e}{ m } \left( \bm{S} \times \bm{B} \right) _{\alpha} +  \frac{e}{2mc^{2}} \epsilon_{jk\alpha} \epsilon_{rlj} E_{l} J^{S}_{rk}=  0, \label{falpha_continuity s-o} \\
&\frac{\partial u_{i} }{\partial t} + u_{j} (\partial_{j} u_{i}) + \frac{1}{nm} \partial_{j} P_{ij} + \frac{e}{m} \left[E_{i} + \left(\bm{\widetilde{u}} \times \bm{B} \right)_{i}  \right]+\frac{e}{nm^{2}}  S_{\alpha} \left(\partial_{i} B_{\alpha} \right) \nonumber \\
&~
+\frac{\mu_B}{2mc^{2} n } \epsilon_{jkl} \left[ u_{i} \partial_{j} \left( E_{k} S_{l} \right) +  E_{j} \left(\partial_{k} J^{S}_{il}\right) -   \left(\partial_{i} E_{k} \right) J^{S}_{jl}   -   \left(\partial_{j} E_{k} \right) J^{S}_{il} \right]
= 0,  \label{euler eq s-o}\\
&\frac{\partial J^{S}_{i \alpha}}{\partial t} + \partial_{j} \Pi_{ij\alpha} + \frac{e E_{i}}{ m }S_{\alpha} + \frac{e}{ m } \epsilon_{jki} B_{k} \widetilde{J}^{S}_{j\alpha} + \frac{e}{ m } \epsilon_{jk\alpha} B_{k} J^{S}_{ij} + \frac{\mu_B \hbar}{2m}  \left(\partial_{i} B_{\alpha} \right) n \nonumber \\
&~
+ \frac{\mu_B}{2mc^{2}} \epsilon_{kl\alpha}\partial_{l} \left( E_{k} n u_{i} \right)- \frac{\mu_B}{2mc^{2} } \epsilon_{kl\alpha}  \left(\partial_{i} E_{l} \right) n u_{k} + \frac{\mu_B}{\hbar c^{2}} \epsilon_{kl\alpha} \epsilon_{rsk} E_{s} \Pi^{S}_{irl} = 0,\label{evol Js s-o}\\
&
\frac{\partial P_{ij}}{\partial t}+ u_{k} \partial_{k} P_{ij} + P_{jk}  \partial_{k} u_{i} + P_{ik} \partial_{k} u_{j}  + P_{ij} \partial_{k} u_{k} +  \partial_{k} Q_{ijk}
+\frac{e}{m} \left[ \epsilon_{kli}  P_{jk}  +  \epsilon_{klj}   P_{ik} \right] B_{l} \nonumber \\
&~
+ \frac{\mu_B}{m}\left[ \partial_{i} B_{k}  \left( J^{S}_{jk} - u_{j} S_{k} \right) + \partial_{j} B_{k}  \left( J^{S}_{ik} - U_{i}S_{k} \right) \right]
+ \frac{\mu_B}{2mc^{2}} \epsilon_{rsl} \partial_{s} \left[ E_{r}\left( \Pi^{S}_{ijl} - u_{i}u_{j}S_{l} \right)\right] \nonumber \\
&~
+  \frac{\mu_B}{2mc^{2}}  \epsilon_{rkp} E_{r} \left[ \epsilon_{kli}   \left(J^{S}_{jp}-u_{j}S_{p} \right) +  \epsilon_{klj}   \left( J^{S}_{ip} - u_{i}S_{p} \right) \right] B_{l}\nonumber \\
&~
- \frac{\mu_B}{2mc^{2}} \epsilon_{rsl} \left[  \partial_{i}  E_{s}  \left(\Pi^{S}_{jrl} - u_{j} J^{S}_{rl} \right)  + \partial_{j}  E_{s}  \left(\Pi^{S}_{irl} - u_{i} J^{S}_{rl} \right)  \right]\nonumber \\
&~
-
\frac{\mu_B}{2mc^{2} }u_{i} \epsilon_{rsl}  \partial_{s} \left[ E_{r} \left( J^{S}_{jl} -  u_{j} S_{l} \right) \right]
-
\frac{\mu_B}{2mc^{2} }u_{j} \epsilon_{rsl}  \partial_{s} \left[ E_{r} \left( J^{S}_{il} -  u_{i} S_{l} \right) \right] =0,
\label{eq_presure s-o}
\end{align}
where we introduced a new average velocity and a new spin current
\begin{align}
 \overline{\bm{u}}  = \bm{u}- \frac{\mu_B}{2mc^{2}n} \bm{E} \times \bm{S},~~~~
\overline{J}^{S}_{ij} = J^{S}_{ij}+ \frac{\mu_B}{2mc^{2}} \epsilon_{ijk} E_{k}n.
\end{align}

Some further hypotheses are needed to close the above set of equations \eqref{f0_continuity s-o}-\eqref{eq_presure s-o}. Inspecting the evolution equation \eqref{eq_presure s-o} for the pressure tensor, one notices that most spin-dependent terms cancel if we set
\begin{align}
J^{S}_{i\alpha} = u_{i} S_{\alpha} ~~~~\textrm{and}~~~~\Pi^{S}_{ij\alpha} = u_{i} J^{S}_{j\alpha}.  \label{int clos s-o}
\end{align}
This is a generalization of the simple closure described in section \ref{Fluid model with spin effects}, whereby both the spin density $S_{\alpha}$ and the spin current $J^{S}_{j\alpha}$ are transported by the mean fluid velocity $u_i$.

With this assumption,  Eq. \eqref{evol Js s-o} and the definition of the spin-pressure $\Pi_{ij\alpha}$ are no longer necessary. The system of fluid equations simplifies to
\begin{align}
&\frac{\partial n}{\partial t} + \bm{\nabla}\cdot \left(n\bm{\overline{u}}\right)   = 0,  \label{f0_continuity closed s-o} \\
&\frac{\partial S_{\alpha}}{\partial t} + \partial_{i}\left(u_{i} S_{\alpha}\right) - \frac{\mu_B}{2mc^{2}} \left( \bm{\nabla} \times n\bm{E} \right)_{\alpha}  + \frac{e}{ m } \left[ \bm{S} \times \left(\bm{B} - \frac{1}{2c^{2}} \bm{u} \times \bm{E} \right) \right] _{\alpha} =  0, \label{falpha_continuity closed s-o} \\
&\frac{\partial u_{i} }{\partial t} + u_{j} (\partial_{j} u_{i}) + \frac{1}{nm} \partial_{j} P_{ij} + \frac{e}{m} \left[E_{i} + \left(\bm{\overline{u}} \times \bm{B} \right)_{i}  \right]+\frac{e}{nm^{2}}  S_{\alpha} \left(\partial_{i} B_{\alpha} \right) \nonumber \\
&~
+\frac{\mu_B}{2mc^{2} n } \epsilon_{jkl} \left[ E_{j} \left(\partial_{k}u_{i}\right) - u_{k} \left( \partial_{i} E_{j} \right)  \right]S_{l}
= 0,  \label{euler eq closed s-o}\\
&
\frac{\partial P_{ij}}{\partial t}+ u_{k} \partial_{k} P_{ij} + P_{jk}  \partial_{k} u_{i} + P_{ik} \partial_{k} u_{j}  + P_{ij} \partial_{k} u_{k} +  \partial_{k} Q_{ijk}
+\frac{e}{m} \left[ \epsilon_{kli}  P_{jk}  +  \epsilon_{klj}   P_{ik} \right] B_{l}=0.
\label{eq_presure closed s-o}
\end{align}
Then, by supposing that the system is isotropic and adiabatic, i.e. $P_{ij} \propto n^{\frac{d+2}{d}} \delta_{ij}$ (where $d$ is the number of degrees of freedom) and $Q_{ijk}=0$, one is able to close the system of fluid equation.

In summary, Eqs. \eqref{f0_continuity closed s-o}-\eqref{euler eq closed s-o}, together with an adapted expression of the pressure, constitute a closed system of hydrodynamic equations including spin-orbit effects.


\section{Variational approach to the quantum fluid models}\label{sec:variational}

\subsection{Basics: Lagrangian formulation of the fluid equations}\label{Principle of the method}

We recall  the quantum fluid equations considered in section \ref{Fluid models}, written here in atomic units ($4\pi\varepsilon_0=e=\hbar=1$):
\begin{align}
&\frac{\partial n}{\partial t} + \bm{\bm{\nabla}}\!\cdot(n\bm{u})=0, \label{eq:continuity_var}\\
&\frac{\partial \bm{u}}{\partial t}+\bm{u}\cdot\!\bm{\bm{\nabla}} \bm{u}= \bm{\bm{\nabla}}\!\,V_H + \bm{\bm{\nabla}}\!\,V_{ ext} -\frac{\bm{\bm{\nabla}} P}{n}+\frac{1}{2}\bm{\bm{\nabla}} \left(\frac{\bm{\bm{\nabla}}^2\sqrt{n}}{\sqrt{n}}\right), \label{eq:momentum}\\
& \bm{\bm{\nabla}}^2 V_H= 4\pi n, \label{eq:poisson_var}
\end{align}
with the degenerate pressure
\(
P={1\over 5}\left(3\pi^2\right)^{2/3}n^{5/3}.
\)

The above fluid equations may be represented by a Lagrangian density $\mathscr{L}(n,\theta,V_H)$, where
$\theta(\bm{r},t)$ is related to the  velocity through $\bm{u} = \bm{\nabla} \theta$. The expression of such Lagrangian density reads as:
\begin{align}
\mathscr{L}(n,\theta,V_H)&=\frac{n}{2}\left(\bm{\bm{\nabla}} \theta\right)^2 +n\frac{\partial \theta}{\partial t}+\frac{1}{8n}\left(\bm{\bm{\nabla}} n \right)^2+
\frac{3}{10}\left(3\pi^2\right)^{2/3}n^{5/3}\nonumber \\
&~
-nV_{ext}-nV_H-\frac{1}{8\pi}\left(\bm{\bm{\nabla}} V_H\right)^2.
\label{eq:lagrange-dens}
\end{align}
By taking the  Euler-Lagrange equations with respect to the three fields $\chi_i = \{n, \theta, V_H\}$
\be
\frac{\partial \mathscr{L}}{\partial \chi_i} - \nabla \cdot \frac{\partial \mathscr{L}}{\partial\nabla \chi_i} -
\frac{\partial}{\partial t} \frac{\partial \mathscr{L}}{\partial \dot{\chi}_i}  = 0 ,
\ee
one recovers exactly the quantum fluid equations \eqref{eq:continuity_var}-\eqref{eq:poisson_var}.

So far, no approximations were made. In order to derive a tractable system of equations, one needs to specify a particular ansatz for the electron density. In general, we posit that the density depends on the position $\bm{r}$ and, parametrically, on  $N$ functions of time $d_{1}(t), \cdots,  d_{N}(t)$, which are not  specified for the moment. Hence we write:
\begin{align}
n(\bm{r},t) = {\cal N}\left[\bm{r}; d_{1}(t), d_{2}(t), \cdots,  d_{N}(t) \right].
\label{density assumption F}
\end{align}
With this assumption, the time dependence of the electronic density is embedded in the dynamical variables $d_{1}(t), \cdots,  d_{N}(t)$. The spatial profile ${\cal N}$ of the density can, for instance, be tuned to match the  ground state of the system.

The reduced Lagrangian $L$ is obtained by integrating the Lagrangian density over all space
\begin{align}
L \left(\{d_{i}(t)\}, \{\dot d_{i}(t)\} \right) = \int \mathscr{L} \left(\bm{r}, \{d_{i}(t)\}, \{\dot d_{i}(t)\}\right) d\bm{r}.
\label{Lagrangian}
\end{align}
Then, the Lagrangian $L$ can be used to find the dynamical equations for the parameters $d_i(t)$ through the Euler-Lagrange equations:
\be
\frac{\partial L}{\partial d_i} - \frac{\textrm{d}}{\textrm{d} t}  \frac{\partial L}{\partial \dot d_i} = 0 ,
\ee
which yield a system of differential equations such as
\begin{align}
\left\{
\begin{array}{ccc}
\ddot{d}_1(t) &=& f_{1}\left(d_{1}(t), \cdots,  d_{N}(t)\right), \\
\ddot{d}_2(t) &=& f_{2}\left(d_{1}(t), \cdots,  d_{N}(t)\right),\\
\vdots && \\
\ddot{d}_N(t) &=& f_{N}\left(d_{1}(t), \cdots,  d_{N}(t)\right).
\end{array}
\right.
\end{align}
In summary, thanks to the above variational approach, the very complex nonlinear many-body dynamics has been translated into a set of ordinary differential equations for the quantities $d_{i}(t)$. The latter may be easily solved with standard numerical methods, such as Runge-Kutta.

For practical applications, the variational method can be difficult to use. Indeed, to find the Lagrangian \eqref{Lagrangian} one has to specify, in addition to the electron density, an analytical expression for the fields $\theta(\bm{r},t)$ and $V_H(\bm{r},t)$ as functions of the dynamical variables. The field $\theta(\bm{r},t)$ is related to the mean velocity and should satisfy the continuity equation
\eqref{eq:continuity_var}, while the field $V_H$ has to obey Poisson's equation \eqref{eq:poisson_var}. It is worth insisting on the fact that we need to find analytical expressions of $\theta$ and $V_{H }$ to make the variational approach useful for applications. However, this can be difficult to do in practice and depends mostly on the mathematical expression that we assumed for the electronic density, i.e. on the function ${\cal N}$ in Eq. \eqref{density assumption F}.
Nevertheless, the procedure generally yields rather robust results. For instance, it was noted in \cite{Haas2009} that, even for an approximate density profile, the resonant frequencies computed with the variational method are still very close to the exact ones, as will be shown in the forthcoming subsection.


\subsection{Application I: Electronic breathing mode in a 1D quantum well}

As an example of application of the variational method, we consider an electron gas  trapped in a 1D harmonic potential
 \footnote{\gm Parts of this section appeared earlier in Ref. \cite{Haas2009}. Copyright (2009) by the American
Physical Society.
}.
This is a situation that can be encountered when modeling the electron dynamics in a semiconductor quantum well.
The evolution of the  electron density $n(x,t)$ and mean velocity $u(x,t)$ is governed by the following 1D
continuity and momentum equations
\begin{eqnarray}
\label{e1}
\frac{\partial n}{\partial t} &+& \frac{\partial}{\partial x}\,(nu) = 0 \,,\\
 \frac{\partial\,u}{\partial\,t} &+& u \frac{\partial
u}{\partial x} = -\frac{1}{\mstar n} \frac{\partial P}{\partial x}
\label{e2} - \frac{1}\mstar\frac{\partial V}{\partial x}
+ \frac{\hbar^2}{2\mstar^2} ~\frac{\partial}{\partial x}
\left(\frac{\partial_x^2 \sqrt{n}}{\sqrt{n}} \right) ,
\end{eqnarray}
where $\mstar$ is the effective electron mass, $\hbar$ is the
reduced Planck constant, $P(x,t)$ is the electron pressure, and
$V=V_{\rm conf}(x)+V_H(x,t)$ is the effective potential,
which is composed of a confining and a Hartree term. The Hartree
potential obeys the Poisson equation, namely $V_H''= -
e^{2}\,n/\varepsilon$, where $\varepsilon$ is the effective dielectric permittivity
of the material.

The pressure $P(x,t)$ in Eq. (\ref{e2}) is related to the
electron density $n$ via a polytropic
equation of state: $P = \overline{n}\,k_B\,T\,(n/\overline{n})^{\gamma}$,
where $\gamma=3$ is the 1D polytropic exponent and $\overline{n}$ is the mean electron
density.

The electrons are confined by a harmonic potential $V_{\rm conf} =
\frac{1}{2}\omega_0^2 \mstar x^2$, where $\omega_0$
can be related to a fictitious homogeneous positive charge of
density $n_0$ via the relation $\omega_0 = (e^2 n_0/\mstar
\epsilon)^{1/2} $. We then normalize time to $\omega_0^{-1}$;
space to $\lambda_{0} = (k_{B}\,T/\mstar)^{1/2}/\omega_{0}$; velocity to
$\lambda_{0}\,\omega_0$; energy to $k_B T$; and the electron
density to $n_0$. Quantum effects are encapsulated in the
scaled Planck constant $H = \hbar\,\omega_{0}/k_{B}\,T$.
These  nondimensional units will be used in the rest of this section.

The  system of Eqs.
(\ref{e1})--(\ref{e2}) corresponds to the following Lagrangian density:
\begin{eqnarray}
{\cal L}(n,\theta,V_H) &=& \frac{1}{2}\left(\frac{\partial V_H}{\partial x}
\right)^2 - n\,V_H - \,n\,\frac{\partial\theta}{\partial t} -
\int^{n}W(n')\,dn' \nonumber \\ \label{e4} &-&
\frac{1}{2}\,\left(n\,\left[\frac{\partial\theta}{\partial
x}\right]^2 + \frac{H^2}{4n}\,\left[\frac{\partial\,n} {\partial
x}\right]^2\right) - n\,V_{\rm conf} \, .
\label{eq:Lagrangian1D}
\end{eqnarray}
The velocity field can be written as the gradient of the ancillary function
$\theta(x,t)$, as $u = \partial\theta/\partial x$. The
potential $W(n)$ in Eq. (\ref{e4}) is a function of the pressure, $W
\equiv \int^{n} \frac{dP}{dn'}\frac{dn'}{n'} =
(3/2)(n/\overline{n})^2$. The Euler-Lagrange equations with respect to $n$,
$\theta$ and $V_H$ yield the equations of motion (\ref{e1})--(\ref{e2}), as
well as the Poisson equation for $V_H$.

We further assume  a Gaussian profile for the electron density
\begin{equation}
\label{e5} n(x,t) = \frac{A}{\sigma}\,\exp\left[-
\frac{(x-d\,)^2}{2\,\sigma^2}\right] \,,
\end{equation}
where $d(t)$ and $\sigma(t)$ are time-dependent functions that
represent the center-of-mass (dipole) and the spatial dispersion
of the electron gas, respectively. The constant $A =
N/\sqrt{2\pi}$, is defined in terms of the total number of electrons $N=\int n\,dx$.

The other fields to be inserted in the Lagrangian \eqref{eq:Lagrangian1D} are
$\theta$ and $V_H$ and they are determined by requiring
that the continuity and Poisson equations are satisfied. Inserting the Ansatz of Eq.
(\ref{e5}) into the continuity equation \eqref{e1}, one obtains: $\theta = (\dot\sigma/2\sigma)(x-d)^2 +
\dot{d}(x-d)$.   The
solution of the Poisson equation with a Gaussian electron density
is
\begin{equation}
\label{e6} V_H = - A\,\sigma\,e^{-\xi^2/2\sigma^2}
 - A\,\sqrt{\frac{\pi}{2}}\,\xi\,{\rm
Erf}\left(\frac{\xi}{\sqrt{2}\,\sigma}\right) + \rm const,
\end{equation}
where ${\rm Erf}$ is the error function.

Finally, one obtains the Lagrangian
\begin{eqnarray}
L & = & \frac{1}{\sqrt{2\pi}\,A}\,\int\,{\cal L}\,dx =
\frac{\dot{d}^2
+ {\dot\sigma}^2}{2} - \frac{d^2 + \sigma^2}{2}  \nonumber \\
&+& \frac{\sqrt{2}}{2}\,A\,\sigma -
\,\frac{\sqrt{3}\,A^2}{6\,\overline{n}^2\,\sigma^2} -
\frac{H^2}{8\,\sigma^2} \,, \label{lag}
\end{eqnarray}
which only depends on  the dipole
$d$ and the variance $\sigma$ and their time derivatives. The Euler-Lagrange equations
corresponding to the Lagrangian $L$ read as
\begin{eqnarray}
\label{e7} \ddot{d} &+& d = 0 , \\
\label{e8} \ddot{\sigma} &+&
\sigma = \frac{\sqrt{2}\,A}{2} +
\frac{\sqrt{3}\,A^2}{3\,\overline{n}^2\,\sigma^3}  +
\frac{H^2}{4\,\sigma^3} \,.
\end{eqnarray}

The confining potential $V_{\rm conf}$ appears in the
harmonic forces on the left-hand side of  Eqs. (\ref{e7}) and
(\ref{e8}). It should be noted that the equations for $d$ and $\sigma$
decouple for purely harmonic confinement. Equation (\ref{e7})
describes rigid oscillations of the electron gas at the effective
plasmonic frequency,  the so-called Kohn mode \cite{Kohn1961}.
Equation (\ref{e8}) describes the dynamics of the breathing mode,
i.e. coherent oscillations of the width of the electron
density around an equilibrium value given by $\ddot{\sigma}= 0$.
The three terms on the right-hand side of Eq. (\ref{e8})
represent respectively  the Coulomb repulsion, the electron
pressure, and the Bohm potential.

\begin{figure}[htb]
\centering
\includegraphics[width=7cm]{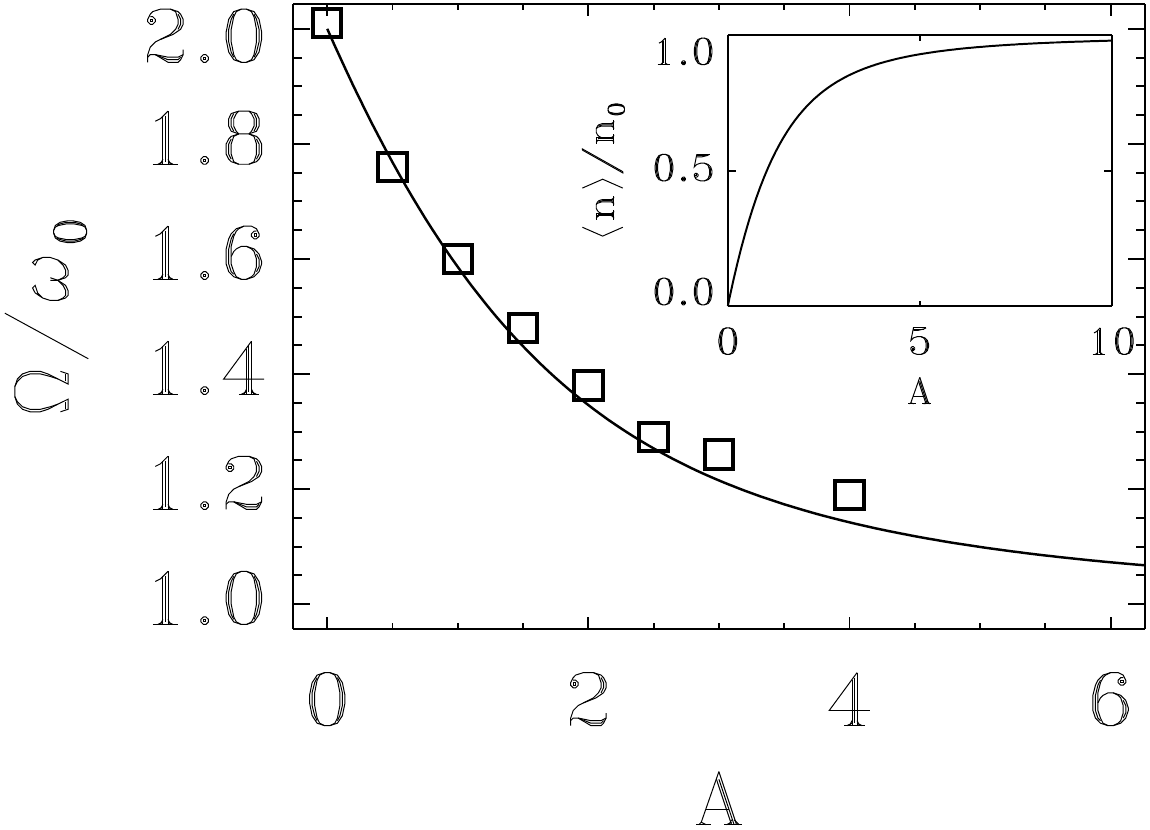}
\caption{The breather frequency $\Omega$ as a function of $A$, for
$H=0.5$. Solid line: analytical results from the variational
method. Squares: The Wigner-Poisson simulations. The inset shows
the mean electron density $\langle n \rangle$
as a function of $A$. Reprinted with permission from \cite{Haas2009}. Copyright (2009) by the American
Physical Society.} \label{fig:breathing}
\end{figure}

The frequency
$\Omega=\Omega(A,H)$ of the breathing mode can be obtained by linearizing the
equation of motion (\ref{e8}) in the vicinity of the stable
fixed point of $U(\sigma)$. The behavior of the breathing
frequency with $A$ (which is proportional to the electron density) is shown in
Fig. \ref{fig:breathing}. For $A=0$ (i.e., when the Coulomb interaction is switched off)
the exact frequency is $\Omega=2\omega_0$. For finite $A$, the
breathing frequency decreases and approaches $\Omega=\omega_0$, for
$A\to\infty$. This is due to the fact that, for large $A$, the electron density becomes flat and equal to $n_0$ (see the inset of
Fig. \ref{fig:breathing}), and is exactly neutralized by the ion density background. For such a spatially homogeneous
system, one can use the results of the Bohm-Gross dispersion relation, which yields $\Omega=\omega_0$ for long
wavelengths.

The above analytical results were compared to numerical simulations of the full Wigner-Poisson equations. The results of the simulations, also plotted in Fig. \ref{fig:breathing}, agree very well with the theoretical curve based on
the Lagrangian approach. The agreement slightly deteriorates for
larger values of $A$, because the electron density deviates from
the Gaussian profile due to strong Coulomb repulsion.

In summary, the variational method described here enabled us to obtain analytically the frequency of the breathing mode oscillations in the linear regime, with excellent agreement in comparison with  ``exact" Wigner-Poisson simulations. Nevertheless, the method is not restricted to the linear response and can be used for large perturbations as well. The only requirement is that the density ansatz \eqref{e5} is approximately verified.

\subsection{Application II: Chaotic electron motion in a 3D nonparabolic anisotropic well}

The same approach was later extended to the case of nonparabolic and anisotropic wells \cite{Hurst2016}.
The confining potential is the sum of a harmonic and an anharmonic (but isotropic) part:
\begin{equation}
V_{\text{conf}}=\frac{1}{2}\left(k_xx^2+k_yy^2+k_zz^2\right) + \zeta \left(x^2+y^2+z^2\right)^{2} \label{eq:Vconf},
\end{equation}
where $k_i>0$ is the stiffness in the $i$-th direction and $\zeta \ge 0$ measures the relative strength of the anharmonic part of the potential.

The algebra is much more convoluted than in the preceding 1D case, but it is still possible to obtain a Lagrangian function $L$  that depends on the six variables $\{d_i(t), \sigma_i(t)\}$ and their time derivatives, where $i=x,y,z$ denotes the Cartesian axes:
\begin{equation}
L[d_i,\sigma_{i},\dot d_i, \dot \sigma_{i}]=\frac{1}{N}\int \mathcal{L} d\bm{r}=\frac{1}{2}\sum_{i}\left(\dot{\sigma_{i}}^{2}+
\dot{d_i}^2\right)- U(d_i,\sigma_i)\label{eq:lagrangian},
\end{equation}
where $N$ is the total number of electrons
and $U(d_{i},\sigma_i)=U_d(d_i)+U_{\sigma}(\sigma_i)+U_{d \sigma}(d_{i},\sigma_i)$. The different potential terms read as:
\begin{align}
U_d=&\frac{1}{2}\sum_{i}k_i d_{i}^{2},\label{eq:Ud}\\
U_{\sigma}=&\nonumber\frac{1}{2}\sum_{i}k_i\sigma_{i}^{2} + \left(\sum_{i}\frac{1}{\sigma_{i}^{2}}\right)\left(\frac{1}{8}+
\alpha_1N\big[\sigma_x\sigma_y\sigma_z\big]^{1/3}\!-
\,\alpha_2\,\beta\bigg[\frac{\sigma_x\sigma_y\sigma_z}{N}\bigg]^{1/3}\right)\label{eq:Usig}\\
&+\,\alpha_3\bigg[\frac{N}{\sigma_x\sigma_y\sigma_z}\bigg]^{2/3}\!\!-
\,\alpha_4\bigg[\frac{N}{\sigma_x\sigma_y\sigma_z}\bigg]^{1/3},\\
U_{d\sigma}=&\zeta\Bigg[\sum_{i}\Big(3\hspace{0.3mm}\sigma_{i}^4+
6\hspace{0.3mm}d_{i}^2\sigma_{i}^2+d_{i}^4\Big)\,+
\sum_{i\ne k}\!\!\Big(\sigma_{i}^2+
d_{i}^2\Big)\Big(\sigma_{k}^2+d_{k}^2\Big)\Bigg]\label{eq:Udsig},
\end{align}
and represent respectively the dipole motion ($U_d$), the breathing motion ($U_{\sigma}$), and the coupling between the dipole and breathing dynamics ($U_{d\sigma}$). Note that such coupling disappears for purely harmonic confinement ($\zeta=0$).
The $\alpha_j$ are  dimensionless coefficients.
The equations of motion obtained from the Euler-Lagrange equations read as:
\begin{align}
&\ddot{d_i}=-\frac{\partial U_{d}}{\partial d_i}-\frac{\partial U_{d\sigma}}{\partial d_i},~~~~~
\ddot{\sigma_i}=-\frac{\partial U_{\sigma}}{\partial \sigma_i}-\frac{\partial U_{d\sigma}}{\partial \sigma_i}. \label{eq:euler-lagrange}
\end{align}

\begin{figure}[htb]
\centering
\includegraphics[width=12cm]{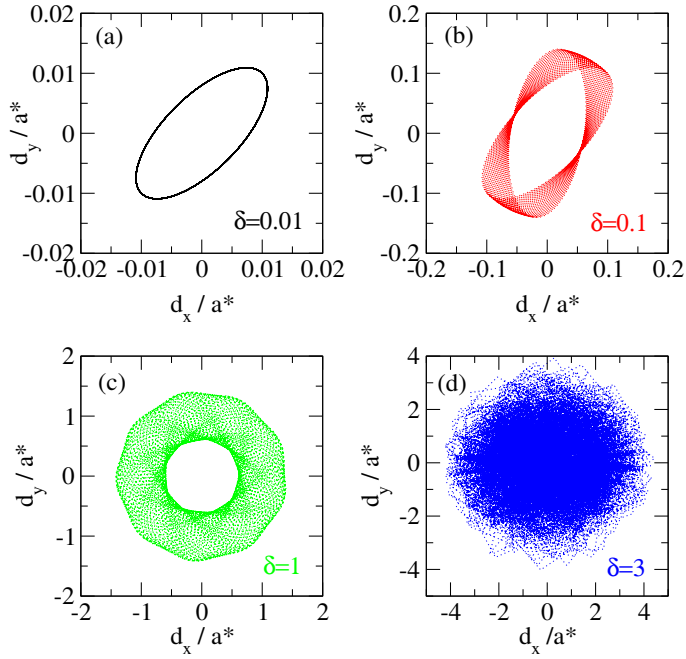}
\caption{Poincar\'e sections in the plane $(d_x, d_y)$ for different values of the initial excitation: $\delta=0.01$ (a), 0.1 (b), 1 (c), and 3 (d). The simulations were performed for an anisotropic well with $k_x=k_y=5$, $k_{z}=1$, $N=50$, and $\zeta=0.01$; $a^{\displaystyle*}$ is a reference length. Reprinted with permission from \cite{Hurst2016}. Copyright (2016) by the American Physical Society.} \label{fig:poincare}
\end{figure}

Figure \ref{fig:poincare} shows the Poincar\'e maps in the $(d_x,d_y)$ plane for increasing values of the initial excitation $\delta$ [$d_i=\delta$, $\dot{d}_{x}=-\dot{d}_{y}=\delta$ and $\dot{d}_{z}=0$, at $t=0$]. At low excitation, the motion is periodic. When $\delta$ is increased, the phase space area progressively fills up, first in a regular way (up to $\delta=1$), and then in an irregular ergodic way. The latter behavior clearly indicates the presence of chaotic motion.


\section{Conclusions and perspectives}
\label{sec:conclusion}

The aim of this short review was to present the results on quantum fluid models obtained in our research group at the University of Strasbourg during the last two decades. We covered the basics of quantum fluid models for systems of spinless particles, then extended the models to include the spin degrees of freedom, as well as semirelativistic corrections. In the last part, we outlined the bases of a variational approach which enables one to reduce the complexity of the fluid models to a set of simple ordinary differential equations for macroscopic quantities such as the center of mass and the width of the electron density.
Several examples taken from past works illustrated the applications of quantum fluid theory to various physical systems, mainly issued from condensed matter and nanophysics. Other areas of possible interest for these methods include laser-plasma interactions \cite{Crouseilles2021} and warm dense matter \cite{Dornheim2018,Fourment2014}. Applications to the astrophysics of compact objects \cite{Uzdensky2014} were not addressed here, but are surely an important  area for future research.

Some care was devoted to establish the validity of the quantum fluid models under consideration (see section \ref{sec:validity}). This was done by comparing the dispersion relations obtained from the fluid models to those derived from a fully kinetic mean-field approach (Vlasov or Wigner). We analyzed in particular the limiting cases of high-frequency (Langmuir) and low-frequency (acoustic) waves. Checking the validity and limits of applicability of the various quantum fluid models is an issue that is unfortunately often neglected in the literature \cite{Bonitz2019}, and we encourage future investigators on these topics to do so systematically.

From a fundamental point of view, the relationships of the quantum fluid methods to the time-dependent density functional theory (TDDFT) \cite{Moldabekov2018,Manfredi2019} and  Thomas-Fermi theory \cite{Michta2015} were highlighted in several past works.
Much less explored is the close link to the so-called orbital-free density functional theory \cite{Witt2018OFDFT}, which appears to be closely related to the present fluid theory. More work  is needed to clarify the similarities and differences between the two approaches.

Finally, some effort should be directed to generalize the quantum fluid models in order to include non-ideal effects, such as dynamical correlations that lead to dissipation and damping. These effects are important on timescales going beyond the electron coherence time, which for standard metallic nano-objects is of the order of 100~fs.
{\gm
In nanophysics,  estimating the viscosity of a quantum electron gas is an open and interesting problem, with applications to graphene and other low dimensional materials \cite{Levitov2016}.
}

There have been some attempts to include potentially dissipative effects in TDDFT. The most accomplished of such attempts is the so-called time-dependent current-density-functional theory  developed by Vignale and Kohn \cite{Vignale97}, which uses the electron current density $\bm{j}(\bm{r},t)$ as the basic building block, instead of the density $n(\bm{r},t)$. However, the relevant equations are mathematically very complicated and not  easy to implement in practical situations.
{\gm
Recently, using the Vignale and Kohn  approach, Cirac\`i  incorporated some viscosity terms in the quantum fluid equations \cite{Ciraci2017}.
}

For phase-space methods, the construction of dissipative terms can rely on the experience acquired in plasma physics, and some examples have been given in our earlier review \cite{Manfredi2019}. How to include such dissipative effects in quantum fluid models, both spinless and spin-dependent, is another important avenue for future research.



\bibliographystyle{spmpsci}
\bibliography{biblio1}

\begin{thebibliography}{10}
\providecommand{\url}[1]{{#1}}
\providecommand{\urlprefix}{URL }
\expandafter\ifx\csname urlstyle\endcsname\relax
  \providecommand{\doi}[1]{DOI~\discretionary{}{}{}#1}\else
  \providecommand{\doi}{DOI~\discretionary{}{}{}\begingroup
  \urlstyle{rm}\Url}\fi

\bibitem{Ali2012}
Al{\`{i}}, G., Mascali, G., Romano, V., Torcasio, R.C.: {A Hydrodynamic Model
  for Covalent Semiconductors with Applications to GaN and SiC}.
\newblock Acta Applicandae Mathematicae  (2012).
\newblock \doi{10.1007/s10440-012-9747-6}.
\newblock \urlprefix\url{http://link.springer.com/10.1007/s10440-012-9747-6}

\bibitem{Andreev2015}
Andreev, P.A.: Separated spin-up and spin-down quantum hydrodynamics of
  degenerated electrons: spin-electron acoustic wave appearance.
\newblock Physical Review E \textbf{91}(3), 033111 (2015)

\bibitem{Anile1995}
Anile, A.M., Muscato, O.: {Improved hydrodynamical model for carrier transport
  in semiconductors}.
\newblock Physical Review B \textbf{51}(23), 16728--16740 (1995).
\newblock \doi{10.1103/PhysRevB.51.16728}.
\newblock \urlprefix\url{http://link.aps.org/doi/10.1103/PhysRevB.51.16728}

\bibitem{Armiento2005}
Armiento, R., Mattsson, A.E.: Functional designed to include surface effects in
  self-consistent density functional theory.
\newblock Phys. Rev. B \textbf{72}, 085108 (2005).
\newblock \doi{10.1103/PhysRevB.72.085108}.
\newblock \urlprefix\url{https://link.aps.org/doi/10.1103/PhysRevB.72.085108}

\bibitem{Arnold1989}
Arnold, A., Steinr\"{u}ck, H.: {The 'electromagnetic' Wigner equation for an
  electron with spin}.
\newblock ZAMP Zeitschrift f\"{u}r angewandte Mathematik und Physik
  \textbf{40}(6), 793--815 (1989).
\newblock \doi{10.1007/BF00945803}.
\newblock \urlprefix\url{http://link.springer.com/10.1007/BF00945803}

\bibitem{Asenjo2011}
Asenjo, F.A., Mu{\~{n}}oz, V., Valdivia, J.A., Mahajan, S.M.: {A hydrodynamical
  model for relativistic spin quantum plasmas}.
\newblock Physics of Plasmas \textbf{18}(1), 012107 (2011).
\newblock \doi{10.1063/1.3533448}.
\newblock
  \urlprefix\url{http://scitation.aip.org/content/aip/journal/pop/18/1/10.1063/1.3533448}

\bibitem{Ashcroft2002}
Ashcroft, N.W., Mermin, N.D., Biet, F., Kachkachi, H., {Impr. Jouve)}:
  {Physique des solides}.
\newblock EDP sciences (2002)

\bibitem{Ciraci2021}
Baghramyan, H.M., Della~Sala, F., Cirac\`{\i}, C.: Laplacian-level quantum
  hydrodynamic theory for plasmonics.
\newblock Phys. Rev. X \textbf{11}, 011049 (2021).
\newblock \doi{10.1103/PhysRevX.11.011049}.
\newblock \urlprefix\url{https://link.aps.org/doi/10.1103/PhysRevX.11.011049}

\bibitem{Banerjee2000}
Banerjee, A., Harbola, M.K.: {Hydrodynamic approach to time-dependent density
  functional theory; Response properties of metal clusters}.
\newblock http://dx.doi.org/10.1063/1.1290610  (2000).
\newblock \doi{10.1063/1.1290610}

\bibitem{Berger2004}
Berger, C., Song, Z., Li, T., Li, X., Ogbazghi, A.Y., Feng, R., Dai, Z.,
  Marchenkov, A.N., Conrad, E.H., First, P.N., et~al.: Ultrathin epitaxial
  graphite: 2d electron gas properties and a route toward graphene-based
  nanoelectronics.
\newblock The Journal of Physical Chemistry B \textbf{108}(52), 19912--19916
  (2004)

\bibitem{Bigot2000}
Bigot, J.Y., Halt{\'{e}}, V., Merle, J.C., Daunois, A.: {Electron dynamics in
  metallic nanoparticles}.
\newblock Chemical Physics \textbf{251}(1), 181--203 (2000).
\newblock \doi{10.1016/S0301-0104(99)00298-0}

\bibitem{Bohm1952}
Bohm, D.: {A Suggested Interpretation of the Quantum Theory in Terms of
  "Hidden" Variables. I}.
\newblock Physical Review \textbf{85}(2), 166--179 (1952).
\newblock \doi{10.1103/PhysRev.85.166}.
\newblock \urlprefix\url{http://link.aps.org/doi/10.1103/PhysRev.85.166}

\bibitem{Bonitz2019}
Bonitz, M., Moldabekov, Z.A., Ramazanov, T.: Quantum hydrodynamics for
  plasmas-quo vadis?
\newblock Physics of Plasmas \textbf{26}(9), 090601 (2019)

\bibitem{Brey1990}
Brey, L., Dempsey, J., Johnson, N.F., Halperin, B.I.: {Infrared optical
  absorption in imperfect parabolic quantum wells}.
\newblock Physical Review B \textbf{42}(2), 1240--1247 (1990).
\newblock \doi{10.1103/PhysRevB.42.1240}.
\newblock \urlprefix\url{http://link.aps.org/doi/10.1103/PhysRevB.42.1240}

\bibitem{Brodin2007}
Brodin, G., Marklund, M.: {Spin magnetohydrodynamics}.
\newblock New Journal of Physics \textbf{9}(8), 277--277 (2007).
\newblock
  \urlprefix\url{http://stacks.iop.org/1367-2630/9/i=8/a=277?key=crossref.225a016bdf04605951bbff2d9f42e59c}

\bibitem{Chan2001}
Chan, G.K.L., Cohen, A.J., Handy, N.C.: {Thomas--Fermi--Dirac--von
  Weizs{\"a}cker models in finite systems}.
\newblock The Journal of Chemical Physics \textbf{114}(2), 631--638 (2001)

\bibitem{Ciraci2017}
Cirac\`{\i}, C.: Current-dependent potential for nonlocal absorption in quantum
  hydrodynamic theory.
\newblock Phys. Rev. B \textbf{95}, 245434 (2017).
\newblock \doi{10.1103/PhysRevB.95.245434}.
\newblock \urlprefix\url{https://link.aps.org/doi/10.1103/PhysRevB.95.245434}

\bibitem{Ciraci2016}
Cirac\`{\i}, C., Della~Sala, F.: Quantum hydrodynamic theory for plasmonics:
  Impact of the electron density tail.
\newblock Phys. Rev. B \textbf{93}, 205405 (2016).
\newblock \doi{10.1103/PhysRevB.93.205405}.
\newblock \urlprefix\url{https://link.aps.org/doi/10.1103/PhysRevB.93.205405}

\bibitem{Ciraci2013}
Cirac\`{\i}, C., Pendry, J.B., Smith, D.R.: Hydrodynamic model for plasmonics:
  A macroscopic approach to a microscopic problem.
\newblock ChemPhysChem \textbf{14}(6), 1109--1116 (2013).
\newblock \doi{https://doi.org/10.1002/cphc.201200992}.
\newblock
  \urlprefix\url{https://chemistry-europe.onlinelibrary.wiley.com/doi/abs/10.1002/cphc.201200992}

\bibitem{Crouseilles2021}
Crouseilles, N., Hervieux, P.A., Li, Y., Manfredi, G., Sun, Y.: Geometric
  particle-in-cell methods for the {V}lasov-{M}axwell equations with spin
  effects.
\newblock Journal of Plasma Physics \textbf{87}(3), 825870301 (2021).
\newblock \doi{10.1017/S0022377821000532}

\bibitem{Crouseilles2008}
Crouseilles, N., Hervieux, P.A., Manfredi, G.: {Quantum hydrodynamic model for
  the nonlinear electron dynamics in thin metal films}.
\newblock Physical Review B \textbf{78}(15), 155412 (2008).
\newblock \doi{10.1103/PhysRevB.78.155412}.
\newblock \urlprefix\url{http://link.aps.org/doi/10.1103/PhysRevB.78.155412}

\bibitem{Domps1998}
Domps, A., Reinhard, P.G., Suraud, E.: {Theoretical Estimation of the
  Importance of Two-Electron Collisions for Relaxation in Metal Clusters}.
\newblock Physical Review Letters \textbf{81}(25), 5524--5527 (1998).
\newblock \doi{10.1103/PhysRevLett.81.5524}.
\newblock \urlprefix\url{http://link.aps.org/doi/10.1103/PhysRevLett.81.5524}

\bibitem{Dornheim2018}
Dornheim, T., Groth, S., Bonitz, M.: The uniform electron gas at warm dense
  matter conditions.
\newblock Physics Reports \textbf{744}, 1--86 (2018).
\newblock \doi{https://doi.org/10.1016/j.physrep.2018.04.001}.
\newblock
  \urlprefix\url{https://www.sciencedirect.com/science/article/pii/S0370157318300516}

\bibitem{Fourment2014}
Fourment, C., Deneuville, F., Descamps, D., Dorchies, F., Petit, S., Peyrusse,
  O., Holst, B., Recoules, V.: {Experimental determination of
  temperature-dependent electron-electron collision frequency in isochorically
  heated warm dense gold}.
\newblock PHYSICAL REVIEW B \textbf{89} (2014).
\newblock \doi{10.1103/PhysRevB.89.161110}

\bibitem{Gunnarsson1976}
Gunnarsson, O., Lundqvist, B.I.: {Exchange and correlation in atoms, molecules,
  and solids by the spin-density-functional formalism}.
\newblock Physical Review B \textbf{13}(10), 4274--4298 (1976).
\newblock \doi{10.1103/PhysRevB.13.4274}.
\newblock \urlprefix\url{http://link.aps.org/doi/10.1103/PhysRevB.13.4274}

\bibitem{Haas2021}
Haas, F.: Exchange fluid model derived from quantum kinetic theory for plasmas.
\newblock Contributions to Plasma Physics \textbf{n/a}(n/a), e202100046.
\newblock \doi{https://doi.org/10.1002/ctpp.202100046}.
\newblock
  \urlprefix\url{https://onlinelibrary.wiley.com/doi/abs/10.1002/ctpp.202100046}

\bibitem{Haas2003}
Haas, F., Garcia, L., Goedert, J., Manfredi, G.: Quantum ion-acoustic waves.
\newblock Physics of Plasmas \textbf{10}(10), 3858--3866 (2003)

\bibitem{Haas2015}
Haas, F., Mahmood, S.: Linear and nonlinear ion-acoustic waves in
  nonrelativistic quantum plasmas with arbitrary degeneracy.
\newblock Physical Review E \textbf{92}(5) (2015).
\newblock \doi{10.1103/physreve.92.053112}.
\newblock \urlprefix\url{http://dx.doi.org/10.1103/PhysRevE.92.053112}

\bibitem{Haas2009}
Haas, F., Manfredi, G., Shukla, P.K., Hervieux, P.A.: {Breather mode in the
  many-electron dynamics of semiconductor quantum wells}.
\newblock Physical Review B \textbf{80}(7), 073301 (2009).
\newblock \doi{10.1103/PhysRevB.80.073301}.
\newblock \urlprefix\url{http://link.aps.org/doi/10.1103/PhysRevB.80.073301}

\bibitem{Haas2010}
Haas, F., Marklund, M., Brodin, G., Zamanian, J.: Fluid moment hierarchy
  equations derived from quantum kinetic theory.
\newblock Physics Letters A \textbf{374}(3), 481--484 (2010).
\newblock \doi{https://doi.org/10.1016/j.physleta.2009.11.011}.
\newblock
  \urlprefix\url{https://www.sciencedirect.com/science/article/pii/S0375960109014248}

\bibitem{Hainfeld2004}
Hainfeld, J.F., Slatkin, D.N., Smilowitz, H.M.: {The use of gold nanoparticles
  to enhance radiotherapy in mice}.
\newblock Phys. Med. Biol. Phys. Med. Biol \textbf{49}(4904), 309--315 (2004).
\newblock \doi{10.1088/0031-9155/49/18/N03}.
\newblock \urlprefix\url{http://iopscience.iop.org/0031-9155/49/18/N03}

\bibitem{Hamann2020}
Hamann, P., Vorberger, J., Dornheim, T., Moldabekov, Z.A., Bonitz, M.: Ab
  initio results for the plasmon dispersion and damping of the warm dense
  electron gas.
\newblock Contributions to Plasma Physics \textbf{60}(10), e202000147 (2020).
\newblock \doi{https://doi.org/10.1002/ctpp.202000147}.
\newblock
  \urlprefix\url{https://onlinelibrary.wiley.com/doi/abs/10.1002/ctpp.202000147}

\bibitem{Hirohata2020}
Hirohata, A., Yamada, K., Nakatani, Y., Prejbeanu, I.L., Di{\'e}ny, B., Pirro,
  P., Hillebrands, B.: Review on spintronics: Principles and device
  applications.
\newblock Journal of Magnetism and Magnetic Materials \textbf{509}, 166711
  (2020)

\bibitem{Hurst2017phd}
Hurst, J.: Ultrafast spin dynamics in ferromagnetic thin films.
\newblock Ph.D. thesis, Universit{\'e} de Strasbourg (2017)

\bibitem{Hurst2017}
Hurst, J., Hervieux, P.A., Manfredi, G.: Phase-space methods for the spin
  dynamics in condensed matter systems.
\newblock Philosophical Transactions of the Royal Society A: Mathematical,
  Physical and Engineering Sciences \textbf{375}, 20160199 (2017).
\newblock \doi{10.1098/rsta.2016.0199}.
\newblock \urlprefix\url{https://doi.org/10.1098/rsta.2016.0199}

\bibitem{Hurst2016}
Hurst, J., L{\'e}v{\^e}que-Simon, K., Hervieux, P.A., Manfredi, G., Haas, F.:
  High-harmonic generation in a quantum electron gas trapped in a nonparabolic
  and anisotropic well.
\newblock Physical Review B \textbf{93}(20), 205402 (2016)

\bibitem{Hurst2014}
Hurst, J., Morandi, O., Manfredi, G., Hervieux, P.A.: {Semiclassical Vlasov and
  fluid models for an electron gas with spin effects}.
\newblock The European Physical Journal D \textbf{68}(6), 176 (2014).
\newblock \doi{10.1140/epjd/e2014-50205-5}.
\newblock \urlprefix\url{http://arxiv.org/abs/1405.1184}

\bibitem{Jones2015}
Jones, R.O.: {Density functional theory: Its origins, rise to prominence, and
  future}.
\newblock Reviews of Modern Physics \textbf{87}(3), 897--923 (2015).
\newblock \doi{10.1103/RevModPhys.87.897}.
\newblock \urlprefix\url{http://link.aps.org/doi/10.1103/RevModPhys.87.897}

\bibitem{Khan1992}
Khan, M.A., Kuznia, J., Van~Hove, J., Pan, N., Carter, J.: Observation of a
  two-dimensional electron gas in low pressure metalorganic chemical vapor
  deposited $\rm gan-al_x ga_{1- x} n$ heterojunctions.
\newblock Applied Physics Letters \textbf{60}(24), 3027--3029 (1992)

\bibitem{Khan2014}
Khan, S.A., Bonitz, M.: Quantum hydrodynamics.
\newblock In: Complex Plasmas, pp. 103--152. Springer (2014)

\bibitem{Klimontovich1960}
Klimontovich, Y.L., Silin, V.P.: { the spectra of systems of interacting
  particles and collective energy losses during passage of charged particles
  through matter}.
\newblock Soviet Physics Uspekhi \textbf{3}(1), 84--114 (1960).
\newblock \doi{10.1070/PU1960v003n01ABEH003260}.
\newblock
  \urlprefix\url{http://stacks.iop.org/0038-5670/3/i=1/a=R04?key=crossref.7bb4e31b1c0804978994fa2e12197175}

\bibitem{Kohn1961}
Kohn, W.: {Cyclotron Resonance and de Haas-van Alphen Oscillations of an
  Interacting Electron Gas}.
\newblock Physical Review \textbf{123}(4), 1242--1244 (1961).
\newblock \doi{10.1103/PhysRev.123.1242}.
\newblock \urlprefix\url{http://link.aps.org/doi/10.1103/PhysRev.123.1242}

\bibitem{Kohn1965}
Kohn, W., Sham, L.J.: {Self-Consistent Equations Including Exchange and
  Correlation Effects}.
\newblock Physical Review \textbf{140}(4A), A1133--A1138 (1965).
\newblock \doi{10.1103/PhysRev.140.A1133}.
\newblock \urlprefix\url{http://link.aps.org/doi/10.1103/PhysRev.140.A1133}

\bibitem{Kremp2001}
Kremp, D., Bornath, T., Hilse, P., Haberland, H., Schlanges, M., Bonitz, M.:
  Quantum kinetic theory of laser plasmas.
\newblock Contributions to Plasma Physics \textbf{41}(2-3), 259--262 (2001)

\bibitem{Levitov2016}
Levitov, L., Falkovich, G.: Electron viscosity, current vortices and negative
  nonlocal resistance in graphene.
\newblock Nature Physics \textbf{12}(7), 672--676 (2016)

\bibitem{Levy1984}
Levy, M., Perdew, J.P., Sahni, V.: Exact differential equation for the density
  and ionization energy of a many-particle system.
\newblock Physical Review A \textbf{30}(5), 2745 (1984)

\bibitem{Lyu2014}
Lyu, L.H.: {Elementary Space Plasma Physics}.
\newblock Airiti Press (2014)

\bibitem{Ma2015}
Ma, W., Miao, T., Zhang, X., Kohno, M., Takata, Y.: Comprehensive study of
  thermal transport and coherent acoustic-phonon wave propagation in thin metal
  film--substrate by applying picosecond laser pump--probe method.
\newblock The Journal of Physical Chemistry C \textbf{119}(9), 5152--5159
  (2015)

\bibitem{Madelung1927}
Madelung, E.: {Quantentheorie in hydrodynamischer Form}.
\newblock Zeitschrift f{\"{u}}r Physik \textbf{40}(3-4), 322--326 (1927).
\newblock \doi{10.1007/BF01400372}

\bibitem{Manfredi2005}
Manfredi, G.: {How to model quantum plasmas}.
\newblock Fields Institute Communications Series \textbf{46}, 263--287 (2005).
\newblock \urlprefix\url{http://arxiv.org/abs/quant-ph/0505004}

\bibitem{Manfredi2001}
Manfredi, G., Haas, F.: {Self-consistent fluid model for a quantum electron
  gas}.
\newblock Physical Review B \textbf{64}(7), 075316 (2001).
\newblock \doi{10.1103/PhysRevB.64.075316}.
\newblock \urlprefix\url{http://link.aps.org/doi/10.1103/PhysRevB.64.075316}

\bibitem{Manfredi2019}
Manfredi, G., Hervieux, P.A., Hurst, J.: Phase-space modeling of solid-state
  plasmas.
\newblock Reviews of Modern Plasma Physics \textbf{3}(1), 1--55 (2019)

\bibitem{Manfredi2018SPIE}
Manfredi, G., Hervieux, P.A., Tanjia, F.: Quantum hydrodynamics for
  nanoplasmonics.
\newblock In: Plasmonics: Design, Materials, Fabrication, Characterization, and
  Applications XVI, vol. 10722, p. 107220B. International Society for Optics
  and Photonics (2018)

\bibitem{Maniyara2019}
Maniyara, R.A., Rodrigo, D., Yu, R., Canet-Ferrer, J., Ghosh, D.S.,
  Yongsunthon, R., Baker, D.E., Rezikyan, A., de~Abajo, F.J.G., Pruneri, V.:
  Tunable plasmons in ultrathin metal films.
\newblock Nature Photonics \textbf{13}(5), 328--333 (2019).
\newblock \doi{10.1038/s41566-019-0366-x}.
\newblock \urlprefix\url{https://doi.org/10.1038/s41566-019-0366-x}

\bibitem{Maurat2009}
Maurat, E., Hervieux, P.A.: Thermal properties of open-shell metal clusters.
\newblock New Journal of Physics \textbf{11}(10), 103031 (2009).
\newblock \doi{10.1088/1367-2630/11/10/103031}

\bibitem{Melrose2020}
Melrose, D.: Quantum kinetic theory for unmagnetized and magnetized plasmas.
\newblock Reviews of Modern Plasma Physics \textbf{4}(1), 1--56 (2020)

\bibitem{Michta2015}
Michta, D., Graziani, F., Bonitz, M.: Quantum hydrodynamics for plasmas -- a
  thomas-fermi theory perspective.
\newblock Contributions to Plasma Physics \textbf{55}(6), 437--443 (2015).
\newblock \doi{https://doi.org/10.1002/ctpp.201500024}.
\newblock
  \urlprefix\url{https://onlinelibrary.wiley.com/doi/abs/10.1002/ctpp.201500024}

\bibitem{Moldabekov2017}
Moldabekov, Z., Bonitz, M., Ramazanov, T.: Gradient correction and bohm
  potential for two- and one-dimensional electron gases at a finite
  temperature.
\newblock Contributions to Plasma Physics \textbf{57}(10), 499--505 (2017).
\newblock \doi{https://doi.org/10.1002/ctpp.201700113}.
\newblock
  \urlprefix\url{https://onlinelibrary.wiley.com/doi/abs/10.1002/ctpp.201700113}

\bibitem{Moldabekov2021b}
Moldabekov, Z., Dornheim, T., B{\"o}hme, M., Vorberger, J., Cangi, A.: The
  relevance of electronic perturbations in the warm dense electron gas.
\newblock arXiv preprint arXiv:2107.00631  (2021)

\bibitem{Moldabekov2021}
Moldabekov, Z., Dornheim, T., B{\"o}hme, M., Vorberger, J., Cangi, A.: The
  relevance of electronic perturbations in the warm dense electron gas.
\newblock arXiv preprint arXiv:2107.00631  (2021)

\bibitem{Moldabekov2018}
Moldabekov, Z.A., Bonitz, M., Ramazanov, T.: Theoretical foundations of quantum
  hydrodynamics for plasmas.
\newblock Physics of Plasmas \textbf{25}(3), 031903 (2018)

\bibitem{Moreau2012}
Moreau, A., Cirac{\`{i}}, C., Mock, J.J., Hill, R.T., Wang, Q., Wiley, B.J.,
  Chilkoti, A., Smith, D.R.: {Controlled-reflectance surfaces with film-coupled
  colloidal nanoantennas}.
\newblock Nature \textbf{492}(7427), 86--89 (2012).
\newblock \doi{10.1038/nature11615}.
\newblock \urlprefix\url{http://www.nature.com/doifinder/10.1038/nature11615}

\bibitem{Muller2004}
M{\"{u}}ller, T., Parz, W., Strasser, G., Unterrainer, K.: {Influence of
  carrier-carrier interaction on time-dependent intersubband absorption in a
  semiconductor quantum well}.
\newblock Physical Review B \textbf{70}(15), 155324 (2004).
\newblock \doi{10.1103/PhysRevB.70.155324}.
\newblock \urlprefix\url{http://link.aps.org/doi/10.1103/PhysRevB.70.155324}

\bibitem{Nie2021}
Nie, Z., Li, F., Morales, F., Patchkovskii, S., Smirnova, O., An, W., Nambu,
  N., Matteo, D., Marsh, K.A., Tsung, F., Mori, W.B., Joshi, C.: In {S}itu
  generation of high-energy spin-polarized electrons in a beam-driven plasma
  wakefield accelerator.
\newblock Phys. Rev. Lett. \textbf{126}, 054801 (2021)

\bibitem{Pines1961}
Pines, D.: Classical and quantum plasmas.
\newblock Journal of Nuclear Energy. Part C, Plasma Physics, Accelerators,
  Thermonuclear Research \textbf{2}(1), 5 (1961)

\bibitem{Romano2001}
Romano, V.: {Non-parabolic band hydrodynamical model of silicon semiconductors
  and simulation of electron devices}.
\newblock Mathematical Methods in the Applied Sciences \textbf{24}(7), 439--471
  (2001).
\newblock \doi{10.1002/mma.220}.
\newblock \urlprefix\url{http://doi.wiley.com/10.1002/mma.220}

\bibitem{Stockman2011}
Stockman, M.I.: {Nanoplasmonics: The physics behind the applications}.
\newblock Physics Today \textbf{64}(2), 39--44 (2011).
\newblock \doi{10.1063/1.3554315}.
\newblock
  \urlprefix\url{http://physicstoday.scitation.org/doi/10.1063/1.3554315}

\bibitem{Tanjia2018}
Tanjia, F., Hurst, J., Hervieux, P.A., Manfredi, G.: Plasmonic breathing modes
  in ${\mathrm{c}}_{60}$ molecules: A quantum hydrodynamic approach.
\newblock Phys. Rev. A \textbf{98}, 043430 (2018).
\newblock \doi{10.1103/PhysRevA.98.043430}

\bibitem{Tans1997}
Tans, S.J., Devoret, M.H., Dai, H., Thess, A., Smalley, R.E., Geerligs, L.,
  Dekker, C.: Individual single-wall carbon nanotubes as quantum wires.
\newblock Nature \textbf{386}(6624), 474--477 (1997)

\bibitem{TatsuroEndo2006}
Tatsuro, E., Kagan, K., Naoki, N., Ha~Minh, H., Do-Kyun, K., Yuji, Y., Koichi,
  N., Eiichi, T.: {Multiple LabelFree Detection of Antigen Antibody Reaction
  Using Localized Surface Plasmon Resonance Based Core Shell Structured
  Nanoparticle Layer Nanochip}  (2006).
\newblock \doi{10.1021/AC0608321}

\bibitem{Thomas1926}
Thomas, L.H.: {The Motion of the Spinning Electron}.
\newblock Nature \textbf{117}(2945), 514--514 (1926).
\newblock \doi{10.1038/117514a0}

\bibitem{Toscano2015}
Toscano, G., Straubel, J., Kwiatkowski, A., Rockstuhl, C., Evers, F., Xu, H.,
  Mortensen, N.A., Wubs, M.: Resonance shifts and spill-out effects in
  self-consistent hydrodynamic nanoplasmonics.
\newblock Nature communications \textbf{6}(1), 1--11 (2015)

\bibitem{Trovato2010}
Trovato, M., Reggiani, L.: {Quantum hydrodynamic models from a maximum entropy
  principle}.
\newblock Journal of Physics A: Mathematical and Theoretical \textbf{43}(10),
  102001 (2010).
\newblock \doi{10.1088/1751-8113/43/10/102001}.
\newblock
  \urlprefix\url{http://stacks.iop.org/1751-8121/43/i=10/a=102001?key=crossref.6b592de46bcaba2db081b3979243c99e}

\bibitem{Tyshetskiy2011}
Tyshetskiy, Y., Vladimirov, S.V., Kompaneets, R.: {On kinetic description of
  electromagnetic processes in a quantum plasma}.
\newblock Physics of Plasmas \textbf{18}(11), 112104 (2011).
\newblock \doi{10.1063/1.3659025}.
\newblock
  \urlprefix\url{http://scitation.aip.org/content/aip/journal/pop/18/11/10.1063/1.3659025}

\bibitem{Uzdensky2014}
Uzdensky, D.A., Rightley, S.: Plasma physics of extreme astrophysical
  environments.
\newblock Reports on Progress in Physics \textbf{77}(3), 036902 (2014)

\bibitem{Vignale97}
Vignale, G., Kohn, W.: Current-dependent exchange-correlation potential for
  dynamical linear response theory.
\newblock Phys. Rev. Lett. \textbf{77}, 2037--2040 (1996).
\newblock \doi{10.1103/PhysRevLett.77.2037}.
\newblock \urlprefix\url{https://link.aps.org/doi/10.1103/PhysRevLett.77.2037}

\bibitem{Vladimirov2011}
Vladimirov, S.V., Tyshetskiy, Y.O.: On description of a collisionless quantum
  plasma.
\newblock Physics-Uspekhi \textbf{54}(12), 1243 (2011)

\bibitem{Voisin2000}
Voisin, C., Christofilos, D., {Del Fatti}, N., Vall{\'{e}}e, F., Pr{\'{e}}vel,
  B., Cottancin, E., Lerm{\'{e}}, J., Pellarin, M., Broyer, M.: {Size-Dependent
  Electron-Electron Interactions in Metal Nanoparticles}.
\newblock Physical Review Letters \textbf{85}(10), 2200--2203 (2000).
\newblock \doi{10.1103/PhysRevLett.85.2200}.
\newblock \urlprefix\url{http://link.aps.org/doi/10.1103/PhysRevLett.85.2200}

\bibitem{Weizsacker1935}
von Weizs{\"a}cker, C.F.: Zur {T}heorie der {K}ernmassen.
\newblock Zeitschrift f{\"u}r Physik \textbf{96}(7-8), 431--458 (1935)

\bibitem{Witt2018OFDFT}
Witt, W.C., Beatriz, G., Dieterich, J.M., Carter, E.A.: Orbital-free density
  functional theory for materials research.
\newblock Journal of Materials Research \textbf{33}(7), 777--795 (2018)

\bibitem{Wu2020}
Wu, Y., Ji, L., Geng, X., Thomas, J., B\"uscher, M., Pukhov, A., H\"utzen, A.,
  Zhang, L., Shen, B., Li, R.: Spin filter for polarized electron acceleration
  in plasma wakefields.
\newblock Phys. Rev. Applied \textbf{13}, 044064 (2020)

\bibitem{Wu2019}
Wu, Y., Ji, L., Geng, X., Yu, Q., Wang, N., Feng, B., Guo, Z., Wang, W., Qin,
  C., Yan, X., et~al.: Polarized electron-beam acceleration driven by vortex
  laser pulse.
\newblock New J. Phys. \textbf{11}, 073052 (2019)

\bibitem{Zamanian2010nj}
Zamanian, J., Marklund, M., Brodin, G.: {Scalar quantum kinetic theory for
  spin-1/2 particles: mean field theory}.
\newblock New Journal of Physics \textbf{12}(4), 043019 (2010).
\newblock \doi{10.1088/1367-2630/12/4/043019}.
\newblock
  \urlprefix\url{http://stacks.iop.org/1367-2630/12/i=4/a=043019?key=crossref.153368f55cfe1c0f5f8e618f46552dfd}

\bibitem{Zamanian2013}
Zamanian, J., Marklund, M., Brodin, G.: Exchange effects in plasmas: The case
  of low-frequency dynamics.
\newblock Phys. Rev. E \textbf{88}, 063105 (2013).
\newblock \doi{10.1103/PhysRevE.88.063105}.
\newblock \urlprefix\url{https://link.aps.org/doi/10.1103/PhysRevE.88.063105}

\bibitem{Zamanian2010}
Zamanian, J., Stefan, M., Marklund, M., Brodin, G.: {From extended phase space
  dynamics to fluid theory}.
\newblock Physics of Plasmas \textbf{17}(10), 102109 (2010).
\newblock \doi{10.1063/1.3496053}.
\newblock
  \urlprefix\url{http://scitation.aip.org/content/aip/journal/pop/17/10/10.1063/1.3496053}

\end{thebibliography}

\end{document}